%% file: main.tex
\documentclass[
aps,
nofootinbib,
reprint,
nobibnotes,
showpacs,
superscriptaddress
]{revtex4-1}

\usepackage[english]{babel}
	\addto\captionsenglish{%
	}

\usepackage{graphicx}

\usepackage{amsmath}
\usepackage{amssymb}
\usepackage[version=3]{mhchem}	
\usepackage{bm}	
\usepackage{nicefrac}	
\usepackage{mathrsfs}
\usepackage{wasysym}

\usepackage[hidelinks]{hyperref}
\hypersetup{
	colorlinks,
	linkcolor={red!70!black},
	citecolor={green!70!black},
	urlcolor={blue!70!black}
}

\usepackage[usenames,dvipsnames]{xcolor}	
\usepackage{comment}	
\usepackage{upgreek}	

\usepackage{textcomp}	

\usepackage[nolist]{acronym}
\usepackage{xspace}

\include{new_commands}

\begin{document}

\title{The CUORE cryostat: an infrastructure for rare event searches at millikelvin temperatures}

\input{0_author_list} 

\date{\today}

\input{0_abstract} 

\maketitle

\input{acronyms}

\input{1_introduction}

\input{2_cryostat_design}

\input{3_vacuum}

\input{4_cool_down}

\input{5_commissioning}

\input{6_CUORE_cool_down}

\input{7_summary}

\input{8_Acknowledgments}

\appendix*

\input{9_glossary}

\section*{} \vspace{-40pt} 

\bibliographystyle{apsrev4-1}
\bibliography{ref_arXiv,ref_Books,ref_Others,ref_Papers,ref_Proceedings,ref_Talks,ref_Theses,ref_web}

\end{document}

%% file: new_commands.tex
\newcommand{\needscitation}[1][{[Citation Needed]}]{\textcolor{red}{#1}\xspace}

		\newcommand{\keV}{\text{keV}}

		\newcommand{\cm}{\text{cm}}	\newcommand{\m}{\text{m}}

\newcommand{\yr}{\text{yr}}

\newcommand{\kg}{\text{kg}}		

	\newcommand{\mK}{\text{mK}}	\newcommand{\K}{\text{K}}

		\newcommand{\W}{\text{W}}

	\newcommand{\bbar}{\text{bar}}

	\newcommand{\s}{\text{s}}			
\newcommand{\Hz}{\text{Hz}}

\newcommand{\ckky}{counts\,$\keV^{-1}\,\kg^{-1}\,\yr^{-1}$}

\newcommand{\mBqkg}{mBq\,$\kg^{-1}$}		\newcommand{\Bqkg}{Bq\,$\kg^{-1}$}		

\newcommand{\mumols}{\upmu\text{mol}\,\s^{-1}}	\newcommand{\mmols}{\text{mmol}\,\s^{-1}}

\renewcommand{\leq}{\leqslant}
\renewcommand{\geq}{\geqslant}

\newcommand{\notaY}{$^{\mbox{\scriptsize $\,\dagger$}}$}

\newcommand{\hdot}{\hphantom{$\cdot$}}

\newcommand{\Jan}{Jan \hspace{2pt}}

\newcommand{\Mar}{Mar \hspace{-.5pt}}

\newcommand{\Jul}{Jul \hspace{4pt}}
\newcommand{\Aug}{Aug \hspace{0pt}}
\newcommand{\Sep}{Sep \hspace{3pt}}
\newcommand{\Oct}{Oct \hspace{2.5pt}}

\newcommand{\sTM}{{\small \texttrademark}} 

\newcommand{\MT}{\mbox{Mini-Tower}}

\newcommand{\yes}{$\checkmark$}%
\newcommand{\no}{$\mathbf \times$}%



%% file: 0_author_list.tex
	\author{C.~Alduino}
	\affiliation{Department of Physics and Astronomy, University of South Carolina, Columbia, SC 29208, USA}

	\author{F.~Alessandria}
	\affiliation{INFN -- Sezione di Milano, Milano I-20133, Italy}

	\author{M.~Balata}
	\affiliation{INFN -- Laboratori Nazionali del Gran Sasso, Assergi (L'Aquila) I-67100, Italy}

	\author{D.~Biare}
	\affiliation{Nuclear Science Division, Lawrence Berkeley National Laboratory, Berkeley, CA 94720, USA}

	\author{M.~Biassoni}
	\affiliation{INFN -- Sezione di Milano Bicocca, Milano I-20126, Italy}

	\author{C.~Bucci}
	\affiliation{INFN -- Laboratori Nazionali del Gran Sasso, Assergi (L'Aquila) I-67100, Italy}

	\author{A.~Caminata}
	\affiliation{INFN -- Sezione di Genova, Genova I-16146, Italy}

	\author{L.~Canonica}
	\altaffiliation[Present address: ]{Max Planck Institut f\"ur Physik, D-80805,  M\"unchen, Germany}
	\affiliation{Massachusetts Institute of Technology, Cambridge, MA 02139, USA}
	\affiliation{INFN -- Laboratori Nazionali del Gran Sasso, Assergi (L'Aquila) I-67100, Italy}

	\author{L.~Cappelli}
	\affiliation{INFN -- Laboratori Nazionali del Gran Sasso, Assergi (L'Aquila) I-67100, Italy}
	\affiliation{Department of Physics, University of California, Berkeley, CA 94720, USA}
	\affiliation{Nuclear Science Division, Lawrence Berkeley National Laboratory, Berkeley, CA 94720, USA}

	\author{G.~Ceruti}
	\affiliation{INFN -- Sezione di Milano Bicocca, Milano I-20126, Italy}

	\author{A.~Chiarini}
	\affiliation{INFN -- Sezione di Bologna, Bologna I-40127, Italy}

	\author{N.~Chott}
	\affiliation{Department of Physics and Astronomy, University of South Carolina, Columbia, SC 29208, USA}

	\author{M.~Clemenza}
	\affiliation{Dipartimento di Fisica, Universit\`{a} di Milano-Bicocca, Milano I-20126, Italy}
	\affiliation{INFN -- Sezione di Milano Bicocca, Milano I-20126, Italy}

	\author{S.~Copello}
	\affiliation{INFN -- Gran Sasso Science Institute, L'Aquila I-67100, Italy}
	\affiliation{INFN -- Laboratori Nazionali del Gran Sasso, Assergi (L'Aquila) I-67100, Italy}

	\author{A.~Corsi}
	\affiliation{INFN -- Laboratori Nazionali del Gran Sasso, Assergi (L'Aquila) I-67100, Italy}

	\author{O.~Cremonesi}
	\affiliation{INFN -- Sezione di Milano Bicocca, Milano I-20126, Italy}

	\author{A.~D'Addabbo}
	\affiliation{INFN -- Laboratori Nazionali del Gran Sasso, Assergi (L'Aquila) I-67100, Italy}

	\author{S.~Dell'Oro}
	\affiliation{Center for Neutrino Physics, Virginia Polytechnic Institute and State University, Blacksburg, Virginia 24061, USA}

	\author{L.~Di Paolo}
	\affiliation{INFN -- Laboratori Nazionali del Gran Sasso, Assergi (L'Aquila) I-67100, Italy}

	\author{M.~L.~Di Vacri}
	\affiliation{INFN -- Laboratori Nazionali del Gran Sasso, Assergi (L'Aquila) I-67100, Italy}
	\affiliation{Dipartimento di Scienze Fisiche e Chimiche, Universit\`a dell'Aquila, L'Aquila I-67100, Italy}

	\author{A.~Drobizhev}
	\affiliation{Department of Physics, University of California, Berkeley, CA 94720, USA}
	\affiliation{Nuclear Science Division, Lawrence Berkeley National Laboratory, Berkeley, CA 94720, USA}

	\author{M.~Faverzani}
	\affiliation{Dipartimento di Fisica, Universit\`{a} di Milano-Bicocca, Milano I-20126, Italy}
	\affiliation{INFN -- Sezione di Milano Bicocca, Milano I-20126, Italy}

	\author{E.~Ferri}
	\affiliation{Dipartimento di Fisica, Universit\`{a} di Milano-Bicocca, Milano I-20126, Italy}
	\affiliation{INFN -- Sezione di Milano Bicocca, Milano I-20126, Italy}

	\author{M.~A.~Franceschi}
	\affiliation{INFN -- Laboratori Nazionali di Frascati, Frascati (Roma) I-00044, Italy}

	\author{R.~Gaigher}
	\affiliation{INFN -- Sezione di Milano Bicocca, Milano I-20126, Italy}

	\author{L.~Gladstone}
	\affiliation{Massachusetts Institute of Technology, Cambridge, MA 02139, USA}

	\author{P.~Gorla}
	\affiliation{INFN -- Laboratori Nazionali del Gran Sasso, Assergi (L'Aquila) I-67100, Italy}

	\author{M.~Guetti}
	\affiliation{INFN -- Laboratori Nazionali del Gran Sasso, Assergi (L'Aquila) I-67100, Italy}

	\author{L.~Ioannucci}
	\affiliation{INFN -- Laboratori Nazionali del Gran Sasso, Assergi (L'Aquila) I-67100, Italy}

	\author{Yu.~G.~Kolomensky}
	\affiliation{Department of Physics, University of California, Berkeley, CA 94720, USA}
	\affiliation{Nuclear Science Division, Lawrence Berkeley National Laboratory, Berkeley, CA 94720, USA}

	\author{C.~Ligi}
	\affiliation{INFN -- Laboratori Nazionali di Frascati, Frascati (Roma) I-00044, Italy}

	\author{L.~Marini}
	\affiliation{Department of Physics, University of California, Berkeley, CA 94720, USA}
	\affiliation{Nuclear Science Division, Lawrence Berkeley National Laboratory, Berkeley, CA 94720, USA}

	\author{T.~Napolitano}
	\affiliation{INFN -- Laboratori Nazionali di Frascati, Frascati (Roma) I-00044, Italy}

	\author{S.~Nisi}
	\affiliation{INFN -- Laboratori Nazionali del Gran Sasso, Assergi (L'Aquila) I-67100, Italy}

	\author{A.~Nucciotti}
	\affiliation{Dipartimento di Fisica, Universit\`{a} di Milano-Bicocca, Milano I-20126, Italy}
	\affiliation{INFN -- Sezione di Milano Bicocca, Milano I-20126, Italy}

	\author{I.~Nutini}
	\affiliation{INFN -- Laboratori Nazionali del Gran Sasso, Assergi (L'Aquila) I-67100, Italy}
	\affiliation{INFN -- Gran Sasso Science Institute, L'Aquila I-67100, Italy}

	\author{T.~O'Donnell}
	\affiliation{Center for Neutrino Physics, Virginia Polytechnic Institute and State University, Blacksburg, Virginia 24061, USA}

	\author{D.~Orlandi}
	\affiliation{INFN -- Laboratori Nazionali del Gran Sasso, Assergi (L'Aquila) I-67100, Italy}

	\author{J.~L.~Ouellet}
	\affiliation{Massachusetts Institute of Technology, Cambridge, MA 02139, USA}

	\author{C.~E.~Pagliarone}
	\affiliation{INFN -- Laboratori Nazionali del Gran Sasso, Assergi (L'Aquila) I-67100, Italy}
	\affiliation{Dipartimento di Ingegneria Civile e Meccanica, Universit\`{a} degli Studi di Cassino e del Lazio Meridionale, Cassino I-03043, Italy}

	\author{L.~Pattavina}
	\affiliation{INFN -- Laboratori Nazionali del Gran Sasso, Assergi (L'Aquila) I-67100, Italy}

	\author{A.~Pelosi}
	\affiliation{INFN -- Sezione di Roma, Roma I-00185, Italy}

	\author{M.~Perego}
	\affiliation{INFN -- Sezione di Milano Bicocca, Milano I-20126, Italy}

	\author{E.~Previtali}
	\affiliation{INFN -- Sezione di Milano Bicocca, Milano I-20126, Italy}

	\author{B.~Romualdi}
	\affiliation{INFN -- Laboratori Nazionali del Gran Sasso, Assergi (L'Aquila) I-67100, Italy}

	\author{A.~Rotilio}
	\affiliation{INFN -- Laboratori Nazionali del Gran Sasso, Assergi (L'Aquila) I-67100, Italy}

	\author{C.~Rusconi}
	\affiliation{Department of Physics and Astronomy, University of South Carolina, Columbia, SC 29208, USA}
	\affiliation{INFN -- Laboratori Nazionali del Gran Sasso, Assergi (L'Aquila) I-67100, Italy}

	\author{D.~Santone}
	\affiliation{INFN -- Laboratori Nazionali del Gran Sasso, Assergi (L'Aquila) I-67100, Italy}
	\affiliation{Dipartimento di Scienze Fisiche e Chimiche, Universit\`{a} dell'Aquila, L'Aquila I-67100, Italy}

	\author{V.~Singh}
	\affiliation{Department of Physics, University of California, Berkeley, CA 94720, USA}

	\author{M.~Sisti}
	\affiliation{Dipartimento di Fisica, Universit\`{a} di Milano-Bicocca, Milano I-20126, Italy}
	\affiliation{INFN -- Sezione di Milano Bicocca, Milano I-20126, Italy}

	\author{L.~Taffarello}
	\affiliation{INFN -- Sezione di Padova, Padova I-35131, Italy}

	\author{E.~Tatananni}
	\affiliation{INFN -- Laboratori Nazionali del Gran Sasso, Assergi (L'Aquila) I-67100, Italy}

	\author{F.~Terranova}
	\affiliation{Dipartimento di Fisica, Universit\`{a} di Milano-Bicocca, Milano I-20126, Italy}
	\affiliation{INFN -- Sezione di Milano Bicocca, Milano I-20126, Italy}
	
	\author{S.~L.~Wagaarachchi}
	\affiliation{Department of Physics, University of California, Berkeley, CA 94720, USA}
	\affiliation{Nuclear Science Division, Lawrence Berkeley National Laboratory, Berkeley, CA 94720, USA}

	\author{J.~Wallig}
	\affiliation{Engineering Division, Lawrence Berkeley National Laboratory, Berkeley, CA 94720, USA}
 
	\author{C.~Zarra}
	\affiliation{INFN -- Laboratori Nazionali del Gran Sasso, Assergi (L'Aquila) I-67100, Italy}

%% file: 0_abstract.tex
\begin{abstract}

	The CUORE experiment is the world's largest bolometric experiment. The detector consists of an array of 988 \ce{TeO_2} crystals, for a total mass of 742\,kg. CUORE is presently 
	taking data at the Laboratori Nazionali del Gran Sasso, Italy, searching for the neutrinoless double beta decay of \ce{^{130}Te}. 
	A large custom cryogen-free cryostat allows reaching and maintaining a base temperature of $\sim 10$\,mK, required for the optimal operation of the detector. 
	This apparatus has been designed in order to achieve a low noise environment, with minimal contribution to the radioactive background for the experiment. 
	In this paper, we present an overview of the CUORE cryostat, together with a description of all its sub-systems, focusing on the solutions identified to satisfy the stringent requirements. 
	We briefly illustrate the various phases of the cryostat commissioning and highlight the relevant steps and milestones achieved each time. 
	Finally, we describe the successful cooldown of CUORE.
	\\[+9pt]
	Published on: Cryogenics {\bf 102}, 9 (2019) \hfill DOI: \href{https://www.sciencedirect.com/science/article/pii/S0011227519301031?via\%3Dihub}{10.1016/j.cryogenics.2019.06.011}

\end{abstract}

%% file: acronyms.tex
\begin{acronym}
\acro{CUORE}{Cryogenic Underground Observatory for Rare Events}
\acro{NH}{Normal Hierarchy}
\acro{IH}{Inverted Hierarchy}
\acro{NDBD}[\ensuremath{0\nu\beta\beta} decay]{Neutrinoless Double Beta Decay}
\acro{LNGS}{Laboratori Nazionali del Gran Sasso}
\acro{HEX}{Heat EXchanger}
\acro{MC}{Mixing Chamber}
\acro{PGA}{Peak Ground Acceleration}
\acro{PT}{Pulse Tube}
\acro{DR}{Dilution Refrigerator}
\acro{DU}{Dilution Unit}
\acro{DCS}{Detector Calibration System}
\acro{OVC}{Outer Vacuum Chamber}
\acro{IVC}{Inner Vacuum Chamber}
\acro{ILS}{Inner Lead Shield}
\acro{MSP}{Main Support Platform}
\acro{TSP}{Tower Support Plate}
\acro{FCS}{Fast Cooling System}
\acro{DS}{Detector Suspension}
\acro{LS}{Lead Suspension}
\acro{ASME}{American Society of Mechanical Engineers}
\acro{WT}{Wire Tray}
\acro{PTFE}{polytetrafluoroethylene}
\acro{OFE}{oxygen-free electrolytic}
\acro{ETP1}{electrolytic tough pitch}
\acro{PGA}{peak-ground-acceleration}
\acro{GM}{Gifford-McMahon}
\acro{NTD}{neutron transmutation doped}
\end{acronym}

\newcommand{\CUORE}{\ac{CUORE}\xspace}
\newcommand{\NH}{\ac{NH}\xspace}
\newcommand{\IH}{\ac{IH}\xspace}
\newcommand{\NDBD}{\ac{NDBD}\xspace}
\newcommand{\LNGS}{\ac{LNGS}\xspace}
\newcommand{\HEX}{\ac{HEX}\xspace}
\newcommand{\MC}{\ac{MC}\xspace}
\newcommand{\PT}{\ac{PT}\xspace}
\newcommand{\DR}{\ac{DR}\xspace}
\newcommand{\DU}{\ac{DU}\xspace}
\newcommand{\DCS}{\ac{DCS}\xspace}
\newcommand{\OVC}{\ac{OVC}\xspace}
\newcommand{\IVC}{\ac{IVC}\xspace}
\newcommand{\ILS}{\ac{ILS}\xspace}
\newcommand{\MSP}{\ac{MSP}\xspace}
\newcommand{\TSP}{\ac{TSP}\xspace}
\newcommand{\FCS}{\ac{FCS}\xspace}
\newcommand{\DS}{\ac{DS}\xspace}
\newcommand{\LS}{\ac{LS}\xspace}
\newcommand{\ASME}{\ac{ASME}\xspace}
\newcommand{\ANSYS}{ANSYS\xspace} 
\newcommand{\WT}{\ac{WT}\xspace}
\newcommand{\PTFE}{\ac{PTFE}\xspace}
\newcommand{\OFE}{\ac{OFE}\xspace}
\newcommand{\ETP}{\ac{ETP1}\xspace}
\newcommand{\PGA}{\ac{PGA}\xspace}
\newcommand{\GM}{\ac{GM}\xspace}
\newcommand{\NTD}{\ac{NTD}\xspace}

%% file: 1_introduction.tex
\section{Introduction}
\label{sec:intro}
	
	The \CUORE is a search for the \NDBD of $^{130}$Te~\cite{Alfonso:2015wka,Alduino:2017ehq} at the \LNGS in Italy. 
	\CUORE uses an array of 988 cryogenic bolometers~\cite{Alduino:2016vjd} and is the largest bolometric experiment in the world. 
	The total active mass of the detector is $742\,\kg$, with about $206\,\kg$ of \ce{^{130}Te}.
	In order to achieve the sensitivity goals of CUORE~\cite{Alduino:2017pni}, the \CUORE cryogenic system must meet a set of very tight design constraints on temperature, operational 
	stability, noise environment, radioactive background and shielding. 
	To achieve these, \CUORE uses a custom-designed cryogen-free cryostat to host and cool the detector down to its operational temperature of $\sim 10\,\mK$.
	
	The CUORE setup is hosted inside a building in Hall A of the \LNGS underground laboratories, where the working volume of the cryostat protrudes into one of the \CUORE cleanrooms. 
	The high cleanliness standards reduced the risk of recontaminating the various parts during the commissioning phase and allowed for safe installation of the detectors.

\begin{figure}[t]
		\centering
		\includegraphics[width=1.\columnwidth]{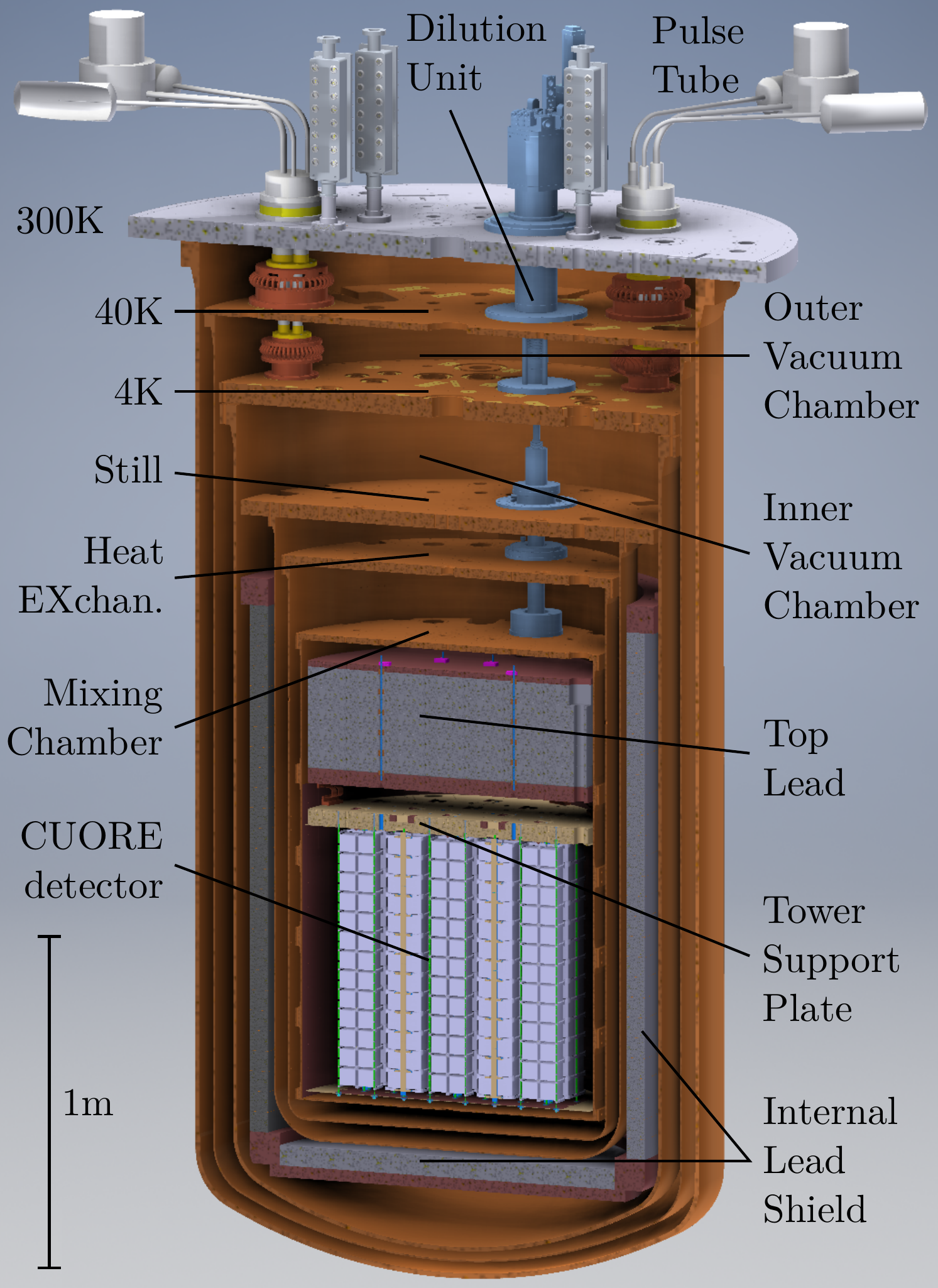}
		\caption{Rendering of the CUORE cryostat. The different thermal stages, vacuum chambers, cooling elements and lead shields are indicated.}
		\label{fig:cryostat_schematic}
	\end{figure}

	A schematic of the CUORE cryostat is shown in Fig.~\ref{fig:cryostat_schematic}.
	The cryostat consists of six nested vessels, the innermost of which contains the experimental volume.
	The different stages step the temperature down from room temperature to $\sim 10\,\mK$, and are identified by their approximate temperatures: 
	300\,K, 40\,K, 4\,K, 800\,mK or Still, 50\,mK or \HEX, and 10\,mK or \MC.

	The 300\,K and the 4\,K vessels are vacuum-tight and define two vacuum volumes called the \OVC and the \IVC, respectively.
	Two lead shields, inside the \IVC, shield the detectors from external radioactivity.
	The \ILS stands between the 4\,K and the Still stages and provides lateral shielding as well as shielding from below. 
	The Top Lead is positioned below the MC plate and provides shielding from above. The \CUORE detector is attached to the \TSP placed right below the Top Lead.

	Cooling is provided through three different systems. The \FCS provides cooling from room temperature during the initial phase of a cooldown. 
	A set of five \PT refrigerators cool the 40\,K and 4\,K plates. The lower stages, along with the detector, are further cooled by a custom-built \ce{^3He/^4He} \DU. 

	The cryostat also embeds the \DCS, which allows deploying radioactive sources for calibrating the detectors (see Ref.~\cite{Cushman:2016cnv} for details).

	In the following sections, we overview the \CUORE cryogenic infrastructure. 
	In Sec.~\ref{sec:cryo_design}, we discuss the mechanical design of the \CUORE cryogenic system, the cryostat and detector support structures, the vibration isolation, 
	and the materials selection. 
	In Sec.~\ref{sec:pressure_vacuum}, we describe the cryogenic vacuum systems. In Sec.~\ref{sec:cool_down}, we detail the cooling subsystems and the cooldown process. 
	In Sec.~\ref{sec:commissioning}, we discuss the commissioning of the \CUORE cryostat and in Sec.~\ref{sec:CUORE_cool_down} we describe the cooldown of the \CUORE detector. 

%% file: 2_cryostat_design.tex
\section{Cryostat design}
\label{sec:cryo_design}

	The design of the \CUORE cryostat represented a significant challenge in the field of the refrigeration technology, coping with a challenging broad range of requirements:
	\begin{itemize}
		\item the experimental volume had to be sufficiently spacious to host the detector and part of the shielding ($\sim 1\,\m^3$);
		\item the detector base temperature had to allow the operation of the bolometers in optimal conditions. 
			From the past experience, the working temperature was $\sim 10\,\mK$~\cite{Andreotti:2010vj,Arnaboldi:2008ds};
		\item the system had to be instrumented with $\sim$2600 twisted pair \ce{NbTi} wires for the detector readout and heater control~\cite{Andreotti:2009zza,Giachero:2013iya};
		\item the background coming from the natural radioactivity of the cryogenic materials had to be compatible with the CUORE sensitivity goal, 
			which demands a total background rate \mbox{$\leq 0.01$\,\ckky} around 2.5\,MeV~\cite{Alduino:2017qet};
		\item the system should be able to perform reliably over a several  year run time with a very high duty cycle to maximize the live-time of the experiment;
		\item since LNGS is located in a seismically active area, the design had to take into account fail-safe mechanisms to limit damage in case of earthquakes.
	\end{itemize}

	The cryostat design began with a conservative thermal budget, supported by numerical modeling, subject to the available refrigeration technologies in 2007-2008, but required several 
	iterations to make sure that enough cooling power was available for any unaccounted for thermal loads. One crucial early decision was the use of a ``dry'' system with multiple 
	{\PT}s, instead of a conventional ``wet'' system which uses a liquid helium bath. 
	This increases the uptime fraction by eliminating the need for He refills, but introduces vibrational challenges. We also chose to use a powerful custom-built \ce{^3He}/\ce{^4He} 
	dilution refrigerator instead of an off-the-shelf dilution unit.

\subsection{Mechanical support}
\label{sec:cryo_support}

	\begin{figure}[tb]
		\centering
		\includegraphics[width=\columnwidth]{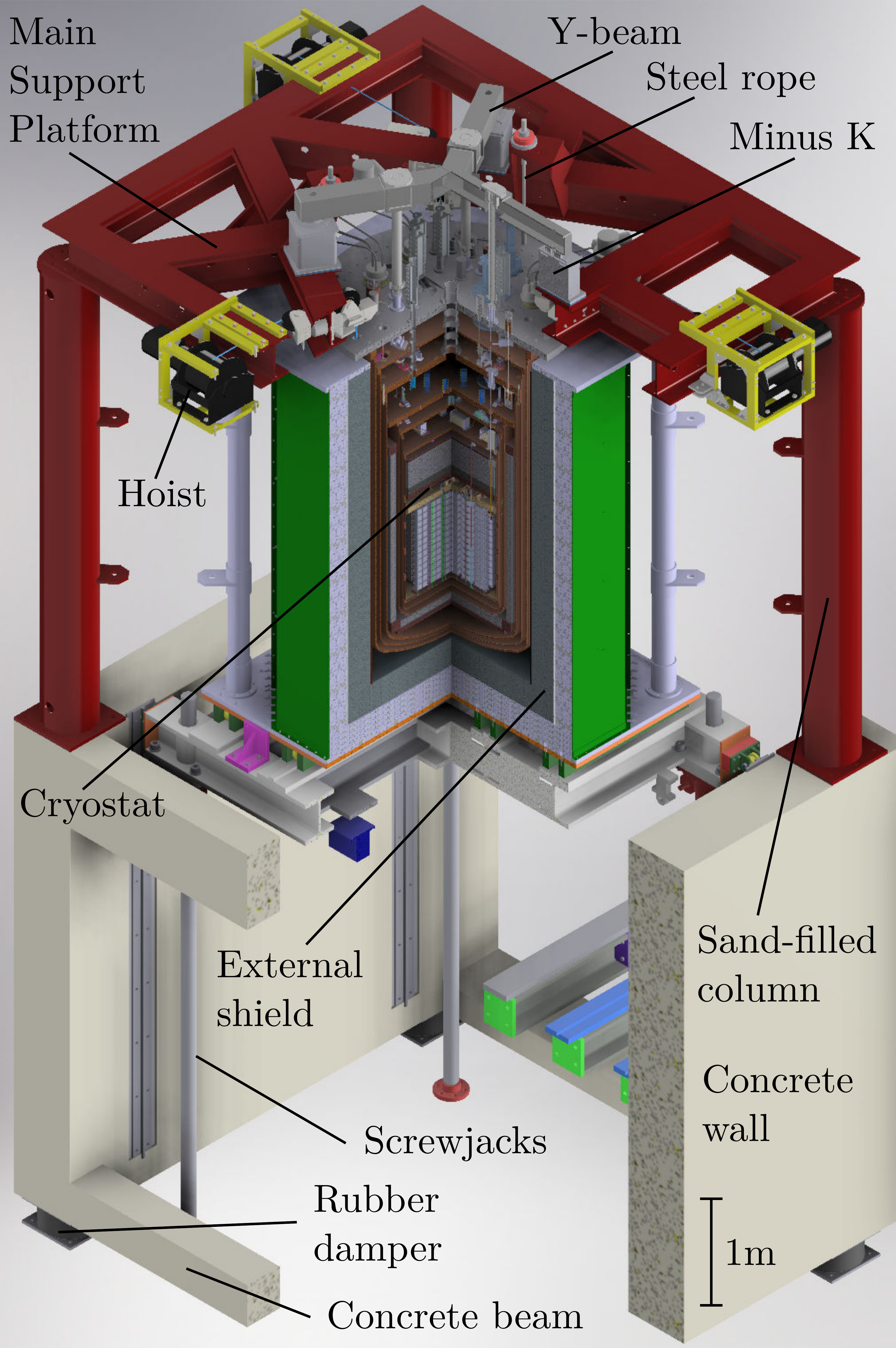}
		\caption{Rendering of the cryostat support structure.  The external lead shield sits on a movable platform and can be raised into place during the detector operation by four screwjacks. 
			When the external lead is lowered, a false floor can close the cryostat and detector into a cleanroom environment.}
		\label{fig:support_structure}
	\end{figure}

	The cryostat support structure (Fig.~\ref{fig:support_structure}) has the double function of being a mechanical support for the cryostat while isolating the cryostat from the 
	rest of the host building and surrounding environment. Further, it must satisfy the earthquake safety requirements for the region. 
	As a result, its design is based on a deep seismic analysis, that has been performed to study the structural response of the system to seismic events~\cite{Ardito:2011SA}.
	
	The lower structure forms the basement and consists of two 60\,\cm~thick walls 4.5\,m tall by 4.5\,m wide connected at the corners by three concrete beams.%
	\footnote{A forth concrete beam was excluded because it would have prevented movement of the external shielding. A still brace (not shown in the figure) was installed at the 
		end of the cryostat commissioning (see Sec.~\ref{sec:commissioning}) to comply with the safety regulations.}
	Each wall rests on two rubber dampers with high damping coefficients, which effectively decouple the structure from the ground.  On top of the walls, four tubular $4.25\,\m$ tall 
	sand-filled steel columns are installed at the corners. The \MSP sits on top of these columns and consists of a $4.7\,\m \times 4.1\,\m$ grid of steel I-beams.

	The 300\,K plate is suspended from the \MSP and sits at the level of the cleanroom ceiling. This plate separates the cleanroom environment, below, 
	from the non-cleanroom environment, above.
	The area above the 300\,K plate and \MSP holds both the plumbing to operate the cryostat as well as the detector front-end electronics inside a Faraday Room~\cite{Bucci:2017gew}.

\begin{figure}[tb]
		\centering
		\includegraphics[width=1.\columnwidth]{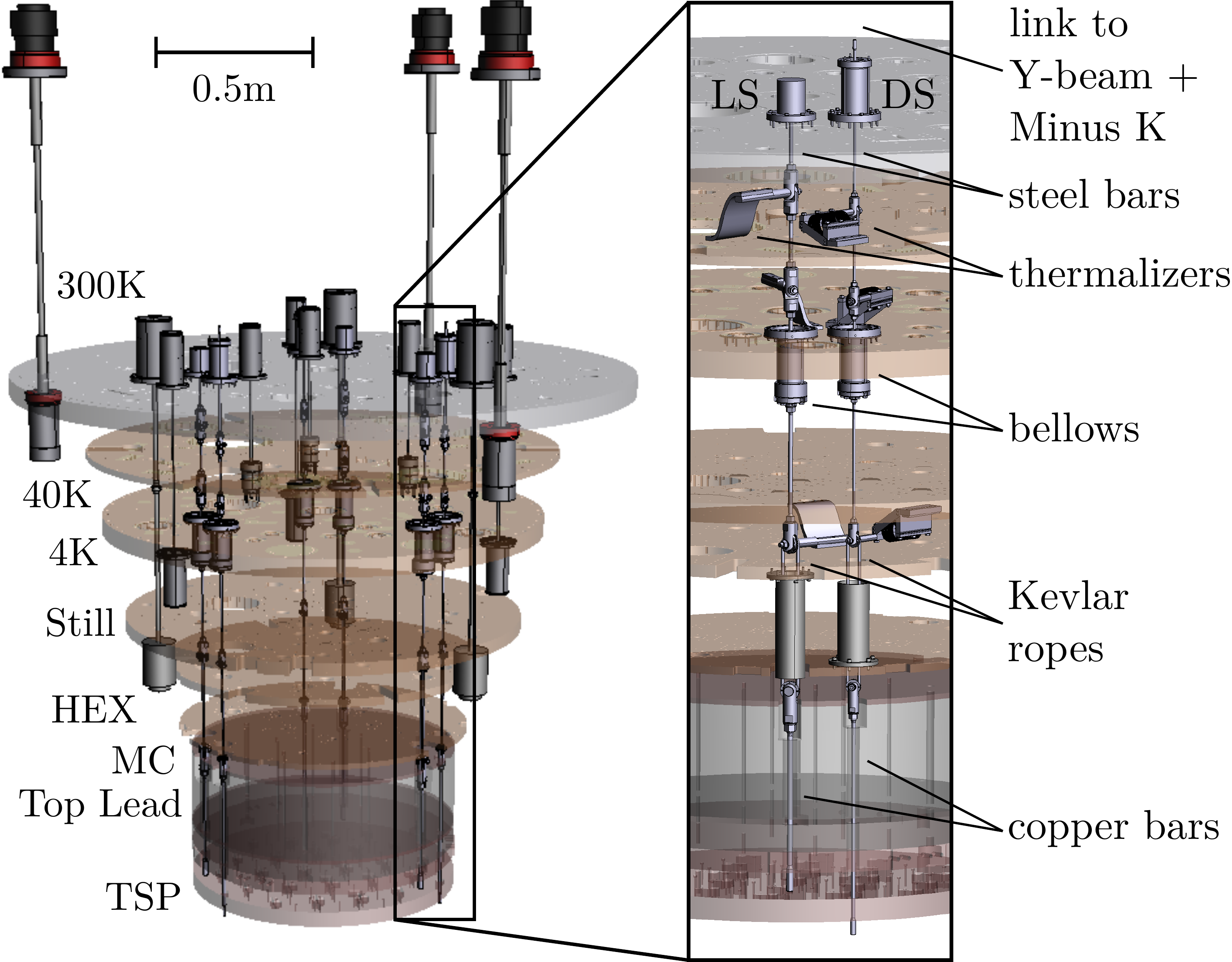}
		\caption{Rendering of open cryostat with detailed view of the supporting elements. The zoom shows the details of the DS and LS components.}
		\label{fig:cryo_structure}
	\end{figure}

	The \MSP supports the full weight of the cryostat by means of three stainless steel ropes.
	Each rope is loaded with $\sim 6$\,t and has been dimensioned to safely withstand the strongest expected earthquake with \PGA $= 0.26\,g$ (the probability of occurrence is once in 475 years).
	All the cryostat components, though not the detector, are directly or indirectly held by the 300\,K plate. 
	Each cryostat plate is held by three vertical bars, that ensure a precise vertical alignment (Table~\ref{tab:bars} and Fig.~\ref{fig:cryo_structure}).
	We performed dedicated measurements to characterize the conductivity of the bars at low temperature and verified that heat leakage through the supports were
	within the thermal budget~\cite{Barucci:2008zz}.

	The 300\,K plate holds the 40\,K, 4\,K and Still plates. The support bars have been designed to be loaded each with about 0.33, 0.66 and 3.2\,t respectively.
	These can resist without damage only more frequent and weaker earthquakes, namely those with \PGA $\sim 0.08\,g$ (the probability of occurrence is once in 35 years).
	The 40 K, 4 K and Still plates are suspended from the OVC top plate by means of three independent sets of 316LN stainless steel bars.
	The bar support are constituted by tilting \ce{CuBe} gimbals designed to prevent damage to the loaded rods in case of a shield displacement caused by a seismic event.%
	\footnote{At the end of the cryostat commissioning phase, the bottom gimbals of the Still support bars have been replaced by brass cylinders in order to reduce the transmission 
		of vibrations, which would have spoiled the detector performance (Sec.~\ref{sec:new_implementations})}.
	In particular, on the 4\,K plate, the vacuum tight access of the Still bars entering the IVC is achieved by metallic bellows welded on the bars themselves.
	The HEX plate is held by the Still plate and supports the MC plate in turn. The support rods are attached to the plates using articulated joints and gimbals to allow 
	relative movement between the individual stages. This prevents structural damage during seismic events.
	
	Each plate holds a corresponding vessel, 
	which encloses the colder volume and acts as thermal radiation shield. Due to the large masses, the vessels are raised and lowered by a system of three synchronized hoists, 
	which are anchored to the \MSP.

	The \ILS is mechanically supported by the Still plate, but thermalized to the 4\,K plate. The Top Lead hangs directly from the 300\,K plate through a set of vertical bars that 
	pass through each of the plates above, but is thermalized to the \HEX plate. 

	The detector is mounted to the \TSP and decoupled from the rest of the cryostat infrastructure.

	\input{9_tab_cryo_structure}

\subsection{Vibration isolation}
\label{sec:mech_requirements}

	Mechanical vibrations generate unwanted heat loads through microfrictions. These can dissipate power on the coldest stages and can prevent stable operation at the desired temperature. 
	The impact is even more severe on the bolometers themselves, which effectively act as heat-to-voltage amplifiers. Mechanical vibrations can impart stray power directly to the 
	detector and this noise, along with microphonic noise, can spoil the energy resolution of the bolometers and directly impact the sensitivity of \CUORE~\cite{Vanzini:2001rx}.

	We adopted various solutions in order to minimize the effects of vibration on the colder stages and detector. 
	The cryostat support structure is designed to mechanically isolate the colder stages from the warmer ones and from the surrounding environment. 
	But this has to be balanced against the need to thermally couple the colder stages to efficiently remove heat and cool them. In particular, the {\PT}s themselves are a significant 
	source of vibration but must be thermally coupled. Hence, a significant effort is made to ensure that vibrational power is dissipated outside the cryostat.

\subsubsection{PT decoupling}
\label{sec:PT_dec}

	In general, {\PT}s have no moving parts at their coldest stages, which significantly reduces the amount of noise generated during operation. 
	However, the compressors and the rotating valves which drive the cooling are sources of mechanical vibrations. 
	Further, the pressure waves generated by compressing and expanding the \ce{He} gas cause a periodic deformation of all the gas lines and induce the cyclical expansion of 
	the actual pulse tube, both along and perpendicularly to the PT axis. Without appropriate countermeasures, 
	we found that the generated vibrational noise would be too high for detector operation.
	
	To minimize the effects of vibrational noise, we start by separating the \PT compressors from the rest of the cryogenic infrastructure.
	The compressors are located $\sim 15$\,m away from the 300\,K plate and are supported by a metallic structure connected to the concrete ground below the CUORE hut via six elastomers. 
	The incoming/outgoing \ce{He} flexlines pass through a sandbox rigidly connected to the ground and filled with $\sim1.5$\,t of non-hygroscopic quartz powder.
	The sandbox absorbs a significant fraction of the vibration generated by the compressors. The high pressure lines are covered with neoprene to reduce the emission of acoustic noise.
	As the lines enter the Faraday Room, a set of ceramic decouplers electrically isolate the lines inside the room from outside and prevent the lines from transmitting 
	electromagnetic noise into the room.

	We separate the rotating valves from the \PT heads (Fig.~\ref{fig:PT}) and mount them separately from the \MSP and 300\,K plate.
	This solution removes all moving parts from the cryostat and thus reduces mechanical vibrations. 
	Moreover, by electrically decoupling the valves from the \PT (and cryostat) we electrically isolate the \PT compressors from the detector grounds. 
	Each \PT head is connected to the 300\,K plate via a polyurethane ring which absorbs part of the incoming vibrations, while a sliding seal compensates for the vertical 
	displacement of the {\PT} due to the thermal contractions at the cold stages (Fig.~\ref{fig:cryostat_schematic}). 
	Despite decoupling from the 300\,K plate, vibration is still transferred down the \PT itself and could be deposited on the lower plates.
	Therefore, the PT cold stages are thermally linked to the 40\,K and 4\,K plates through a set of flexible thermal links (copper braids), 
	making a softer connection and reducing vibration transfer to the plates themselves.

	Separating the rotating valves from the \PT head also mitigates the effects of the \ce{He} pressure waves, by allowing all contraction and expansion to be absorbed by movement 
	of the valve itself. The rotating valves and gas expansion vessels are attached to the compressors through a set of custom high pressure lines which run along the ceiling of the 
	Faraday room. The pressure lines and rotating valves are supported by a set of bungee cords anchored to the ceiling of the Faraday Room, allowing constrained movement of the 
	valves and lines. This design dissipates vibrations into the cords and the Faraday Room, and away from the \MSP and 300\,K plates.

	Finally, instead of using the standard stepping motor embedded in the \PT compressors, the \CUORE rotating valves are driven by microstepping motors specifically intended 
	for situations where motor vibrations must be minimized. These custom drivers also allow for control over the relative phases of the running \PT pressure oscillations and 
	tune them to a configuration that minimizes vibration transmission to the detector~\cite{DAddabbo:2017efe}.

\subsubsection{Detector decoupling}
\label{sec:detector_dec}

	In order to further reduce vibration transfer from the cooling infrastructure, the \CUORE detector is mechanically decoupled from the cryostat. 
	The detector, mounted to the \TSP, hangs from a steel Y-beam positioned on top of three Minus\,K vibration isolators~\cite{MinusK} directly anchored to 
	the MSP (Fig.~\ref{fig:support_structure}). 
	Each isolator behaves like a soft spring when subjected to small displacements, despite the heavy load it bears ($\sim 1$ tonne in total). 
	This behavior favors very low natural frequencies for the spring-mass system, with a cut-off frequency of close to $0.5\,\Hz$ (see Ref.~\cite{Bersani:DS_prep} for details). 

	The connection between the TSP and the Y-beam is made by means of the \DS system.
	The \DS system, as well as the \LS system that supports the Top Lead, consists of three segmented vertical bars (Fig.~\ref{fig:cryo_structure} and Table~\ref{tab:bars}). 
	The top part, from the Y-beam (or 300\,K plate for the \LS) down to the Still level, is made of segmented steel rods, which are thermalized at each cold stage and pass into 
	the \OVC and \IVC through a set of vacuum-tight bellows. 
	From the Still down to through the \MC each support consist of a pair of Kevlar ropes, which minimize heat transfer down to the coldest stages.%
	\footnote{The cryogenic properties of Kevlar K49 have been studied in a dedicated measurement, focusing on the long-term mechanical creeping~\cite{Bersani:2013zta}.} 
	Finally, the \TSP is supported from the bottom of the Kevlar ropes through a set of high purity copper rods, chosen for their low radioactivity content.

\subsection{Materials}
\label{sec:materials}

	\input{9_tab_cryo_masses}

	One of the main goals of the \CUORE cryostat is to create a low background environment for the \CUORE detector to search for \NDBD and other rare events. 
	This places strong constraints on the materials chosen to build the various components of the cryogenic infrastructure, as well as on their handling and storage. 
	A major challenge was to select materials which had minimal bulk \ce{Th} and \ce{U} contamination, while still being compatible with the low temperature thermal requirements.
	We carried out an extensive campaign of radio-assaying to select materials and production techniques for all components. 
	The resulting activity measurements and upper limits are used as input to a detailed Monte Carlo simulation to evaluate the expected background for \CUORE~\cite{Alduino:2017qet}.

	The majority of the structural cryostat components, including the plates and vessels, are made from copper (Table~\ref{tab:mass_inventory}). 
	A significant exception is the 300\,K plate and the upper part of the 300\,K vessel flange, which are made from austenitic stainless steel. 
	The use of stainless steel provides a more reliable vacuum seal and more mechanical strength against the huge load across the area of the plate. 
	The lower plates and vessels down to the \HEX, as well as the lower part of the 300\,K vessel are made from \OFE C10100 copper, which has a purity of 99.99\%.
	Assays of the \CUORE \OFE copper have placed upper limits on the \ce{^{232}Th} and \ce{^{238}U} activity at $<6.5\cdot 10^{-5}$\Bqkg~and $<5.4\cdot10^{-5}$\Bqkg, respectively.%
	\footnote{A detailed description of the vessel production and machining can be found in Ref.~\cite{Alessandria:2013ufa},  where special attention is paid to the description of 
		the welding techniques employed to minimize radioactive contamination during welding.}

	The \MC vessel, \MC plate, and the \TSP are made of electrolytic tough pitch copper, referred to here as NOSV copper~\cite{CuNOSV_aurubis}. 
	This is the same copper used to produce the frames for the \CUORE detector. This type of copper was selected for its high thermal conductivity at low temperatures 
	(Residual-resistance ratio $RRR \geq 400$), 
	its low radioactivity, and for its low hydrogen content.
	The latter property is important since, at low temperatures, the slow conversion of ortho-hydrogen to para-hydrogen can create heat leaks, 
	dissipating power on the coldest stages~\cite{Schwark&al:1983,Kolac&al:1984}.
	A dedicated measurement of the heat leak by NOSV copper set an upper limit of $3.7$\,pW\,g$^{-1}$, well below usual the value for standard copper~\cite{Martinez:2009gjv}.
	Additionally, as the NOSV copper is some of the closest material to the \CUORE detector, the limit on its radioactivity is even more stringent than on the \OFE copper. 
	After procurement, all copper has been stored underground at \LNGS to minimize activation by cosmic radiation.

	The lead shielding, inserted to protect the detector from the external radiation, underwent the material selection process as well.
	The \ILS~\cite{Bucci:ILS_prep} is a $6\,\cm$-thick lead vessel which shields the detector from the sides and bottom. 
	The choice of lead for the \ILS was critical, as this shield is only separated from the \CUORE detector by $\sim1\,\cm$ of copper (i.\,e.\ the thickness of the HEX and MC vessel). 
	In particular, the presence of \ce{^{210}Pb} creates a low energy background continuum in \CUORE through the production of bremsstrahlung produced in its $\beta$-decay and 
	that of the daughter \ce{^{210}Bi}. 
	Because of this, the \ILS is made of radio-pure lead of archaeological origin, with an upper limit on its \ce{^{210}Pb} bulk contamination of 4\,\mBqkg~\cite{Alessandrello:1998RL}. 
	The Top Lead consists of five $6\,\cm$-thick disks of commercial pure lead stacked and sandwiched between two copper plates. 
	Here, the choice of lead was less critical since the Top Lead is separated from the detectors by over $9\,\cm$ of copper (i.\,e.\ the bottom plate of the Top Lead + the TSP). 
	The \ILS supports are made from high purity \OFE copper, and the copper plates sandwiching the Top Lead are made from NOSV copper.

	In order to reduce the amount of heat transferred from the warmer stages to the colder stages as radiation, we cover the vessels and plates of the 40\,K and 4\,K stages 
	in Coolcat\,2\,NW thermal insulation by Ruag (referred to here as superinsulation). 
	Each layer of superinsulation is composed of 10 double-sided aluminized, perforated polyester foil interleaved with 10 foils of non-woven polyester spacer material. 
	The 40\,K stage is wrapped with three such layers, while the 4\,K stage is wrapped in one layer. Given the large surface area, 
	this actually represents a significant mass of superinsulation of $\sim17\,\kg$. 
	However, despite its large mass, the superinsulation contributes negligibly to the radioactive background of the detectors~\cite{Alduino:2017qet}.

%% file: 9_tab_cryo_structure.tex
	\begin{table}[tb]
		\small
		\caption{Details of the cryostat support structure. For each plate and shield, the mechanical mount point, the distance to the above stage and the characteristics of the 
			support bars (material and geometry) are shown. The Top Lead and \TSP have multiple support stages, which are listed from top to bottom. See also Fig.~\ref{fig:cryo_structure}.}
		\begin{ruledtabular}
		\begin{tabular}{l r r r r r}
		Plate		&Mount				&Dist.			&Material		&Diam.			&Joints				\\
					& Point					&[mm]				&					&[mm]				&						\\
		\hline
		300\,K	&MSP				&750				&steel			&$22$				&steel				\\
					&					&					&					&					&joints				\\[+3pt]

		40\,K		&300\,K			&$195$			&SS 316LN		&$8$				&CuBe					\\
					&					&					&					&					&gimbals				\\[+3pt]

		4\,K		&300\,K			&$198$			&SS 316LN		&$7.5$			&CuBe					\\
					&					&					&					&					&gimbals				\\[+3pt]

		Still		&300\,K			&$287$			&SS 316LN		&$18$				&CuBe					\\
					&					&					&					&					&gimbals	\\[+3pt]

		HEX		&Still			&$112$			&Ti6Al4V			&$6$				&steel				\\
					&					&					&					&					&joints				\\[+3pt]

		MC			&HEX				&$202$			&Ti6Al4V			&$4$				&steel				\\
					&					&					&					&					&joints				\\[+3pt]

		ILS		&Still			&$187$			&steel			&$20$				&steel				\\
					&					&					&					&					&threaded			\\
					&					&					&					&					&bushings			\\[+3pt]

		Top Lead	&300\,K\notaY	&$60$				&SS 316LN		&$8.7$			&steel /				\\
					&Still			&					&Kevlar\,K49	&$8\times2$		&copper				\\
					&MC				&					&Cu NOSV			&$12.2$			&blocks				\\[+3pt]

		TSP		&Y-beam			&$76.8$			&SS 316LN		&$12$				&steel /				\\
					&300\,K\notaY	&					&SS 316LN		&$6$				&copper				\\
					&Still			&					&Kevlar\,K49	&$6\times2$		&block				\\
					&MC				&					&Cu NOSV			&$9.3$			&						\\
		\end{tabular}
		\\[+8pt]
		\begin{flushleft}
		\footnotesize
		\notaY segmented: 300\,K\,--\,\,40\,K\,/\,40\,K\,--\,\,4\,K\,/\,4\,K\,--\,\,4\,K\,/\,4\,K\,--\,\,Still \\
		\end{flushleft}
		\end{ruledtabular}
		\label{tab:bars}
	\end{table}

%% file: 9_tab_cryo_masses.tex
	\begin{table}[tb]
		\centering
		\small
		\caption{Cryostat mass inventory.}
		\begin{ruledtabular}
		\begin{tabular}{l r r r r r}
			Part				&Diam.			&Height			&Thick.		&Material		&Mass		\\
								&[mm]				&[mm]				&[mm]			&					&[kg]		\\
		\hline
		\\[-7pt]
			\emph{Vessels} \\				\cline{1-1} \\[-7pt]
			300\,K plate	&2060				&62				&-				&SS 304L		&1540		\\
			300\,K shield	&1603				&2620				&12/15		&Cu OFE 		&1600		\\
								&					&410				&12			&+ SS 304L	&370		\\[+3pt]	

			40\,K plate		&1573				&20				&-				&Cu OFE		&300		\\		
			40\,K shield	&1503				&2765				&5				&Cu OFE		&680		\\[+3pt]

			4\,K plate		&1473				&59				&-				&Cu OFE		&830		\\		
			4\,K shield		&1363				&2471				&10/12		&Cu OFE		&1160		\\[+3pt]

			Still plate		&1333				&43				&-				&Cu OFE		&500		\\		
			Still shield	&1110				&1850				&5				&Cu OFE		&340		\\[+3pt]

			HEX plate		&1070				&28				&-				&Cu OFE		&220		\\		
			HEX shield		&1020				&1650				&5				&Cu OFE		&290		\\[+3pt]

			MC plate			&980				&18				&-				&Cu NOSV		&110		\\		
			MC shield		&940				&1365				&5				&Cu NOSV		&340		\\[+3pt]

		\\[-7pt]
		\emph{Lead shields} \\		\cline{1-1} \\[-7pt]
			ILS (side)		&1157				&1568				&60			&Pb {\footnotesize (Roman)}	&3920		\\		
			ILS (bottom)	&1080				&60				&-				&Pb {\footnotesize (Roman)}	&620		\\		
			Ring (top)		&1157				&80				&88			&Cu OFE		&240		\\
			Ring (bottom)	&1080				&95				&98.5			&Cu OFE		&300		\\
			Plate (bottom)	&1150				&35				&-				&Cu OFE		&320		\\[+3pt]

			Top Lead			&900				&300				&-				&Pb			&2110		\\
			Plate (top)		&900				&18				&-				&Cu NOSV		&90		\\
			Plate (bottom)	&900				&46				&-				&Cu NOSV		&260		\\[+3pt]

		\\[-7pt]
		\emph{Detector} \\			\cline{1-1} \\[-7pt]
			TSP				&900				&48				&-				&Cu NOSV		&260		\\
			Frames			&-					&-					&-				&Cu NOSV		&67		\\
			Crystals			&-					&-					&-				&\ce{TeO_2}	&742		\\[+3pt]		
		\end{tabular}
		\end{ruledtabular}
		\label{tab:mass_inventory}
	\end{table}

%% file: 3_vacuum.tex
\section{Vacuum systems}
\label{sec:pressure_vacuum}

	\begin{table}[tb]
		\begin{center}
		\caption[OVC and IVC maximum differential pressures (table)]{Maximum design differential pressures (inside vs.\ outside and outside vs.\ inside) 
			for the OVC and IVC vessels that can be stood by the volumes during the different phases of operations.}
		\begin{ruledtabular}
		\begin{tabular}{l rrr}
		Vessel		&V $[\m^3]$	&\multicolumn{2}{c}{$\Delta P_{\max}\,[\text{bar}]$}	\\[+2pt]
						\cline{3-4} \\[-12pt]
				      &				&$\Delta P_\mathrm{in-out}$	&$\Delta P_\mathrm{out-in}$	\\
		\hline
		OVC			&5.86			&0.1			&1.01						\\
		IVC			&3.44			&1.5			&1.1\hphantom{0}		\\
		\end{tabular}
		\end{ruledtabular}
		\label{tab:pressures}
		\end{center}
	\end{table}

	The \CUORE cryostat has two separate vacuum chambers: the \OVC, which comprises the 300\,K vessel and the \IVC, which comprises the 4\,K vessel. 
	The two nested vacuum chambers are essential for the cooldown process, which uses \ce{He} exchange gas in the \IVC to thermalize the colder cryostat stages.

	The 300\,K and the 4\,K vessels must be able to withstand both internal and external overpressure without worsening the vacuum and avoiding plastic deformations.
	While the OVC is subject to ambient pressure, the IVC has to withstand $\sim 1.3\,\bbar$ of internal overpressure during the initial phase of the cooldown, when the \FCS 
	is operational (see Sec.~\ref{sec:FCS}).

	The wall thickness for the vacuum vessels were calculated according to the \ASME code~\cite{ASME-BPVC} and we performed a linear buckling analysis with \ANSYS~\cite{ANSYS-web} 
	in order to ensure that we could reach the pressure requirements (Table~\ref{tab:pressures}).
	The limiting differential pressures for the \OVC and \IVC vessels have been calculated to be 2.0\,bar and 2.3\,bar, respectively. A drop-off plate on the \OVC and a rupture disk on the 
	\IVC prevent the internal pressure of each vessel from exceeding the environment pressure by $0.5\,\bbar$.

	The pumping system consists of one turbomolecular pump for each volume backed by a single dry scroll pump. 
	In particular, the \IVC turbomolecular pump was selected for its elevated compression factor for \ce{He} ($>10^8$), which minimizes the presence of residual gas after the completion of 
	the \FCS phase of the cooldown. A charcoal getter (with integrated heater) placed below the 4\,K plate inside the \IVC helps in absorbing any \ce{He} leftover.%
	\footnote{The charcoal getter was inserted at the end of the cryostat commissioning since its presence makes it difficult to perform leak tests.}
	
	The seal between the 300\,K plate and vessel is made with a Buna-N elastomer ring, selected for the low \ce{He} permeability. For the \IVC, the seal must hold at cryogenic 
	temperatures and thus the connection between the 4\,K plate and vessel is made by a Helicoflex\sTM~seal~\cite{Helicoflex_technetics} mounted inside a centering ring.
	In fact, elastomer o-rings cannot be used at cryogenic temperatures since they would freeze, while the use of both indium and Kapton sealings would be impractical given the 
	dimension of the flange.

	Elastomer o-rings are used to seal the various OVC and IVC ports at room temperature, while indium gaskets are used for the IVC ports on the 4\,K plate.

\begin{figure}[tb]
	\centering
	\includegraphics[width=1.\columnwidth]{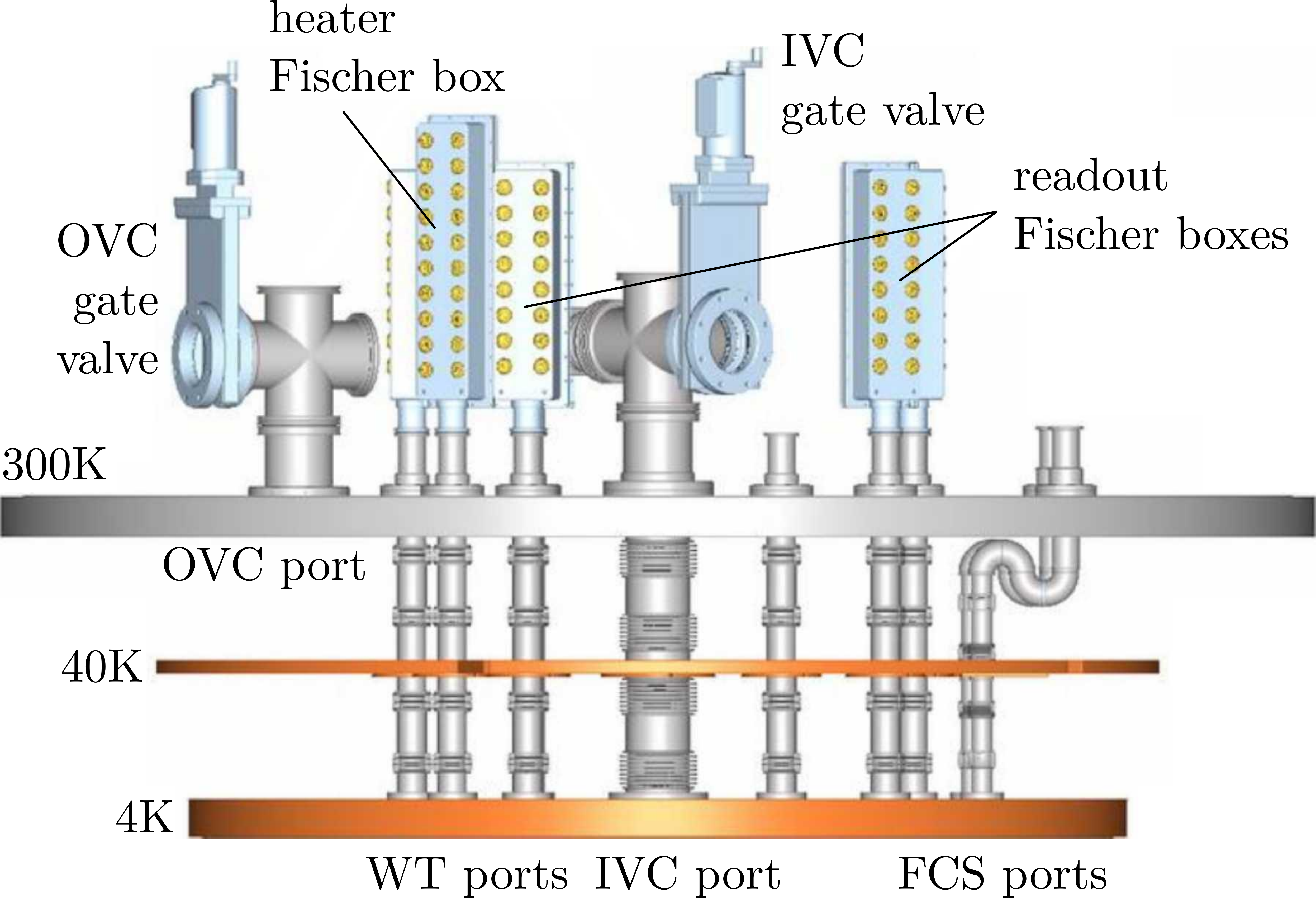}
		\caption{Schematic of the OVC and IVC ports. The WT, heater and diagnostic sensor ports are connected to Fischer boxes. 
			The OVC and IVC ports are closed gate valves for the pumping and the injection of gas. The FCS ports are sealed with blank flanges when the system is not in operation. 
			The lines inside the cryostat are provided with bellows, which allows for the system flexibility in case of seismic events.}
	\label{fig:access_ports}
	\end{figure}

	The \OVC has 2 access ports (Fig.~\ref{fig:access_ports}): 1 DN100 pumping port, and 1 DN40 wiring port for the readout of temperature and pressure sensors.
	The \IVC has multiple access ports: 1 DN100 pumping port, 5 DN40 \WT ports for the detector readout, 1 DN40 \WT port for heater control, 1 DN40 \WT port for readout of 
	temperature and pressure sensors, and 2 DN40 ports for the inlet/outlet of the \FCS \ce{He}. Furthermore, there are 4 DN50 ports for the deployment of the radioactive sources 
	with the \DCS and 3~DN16 ports for the readout of the \DU sensors embedded in the \DU.

	All the \IVC ports are equipped with radiation baffles that prevent direct line of sight between the 300\,K and the 4\,K plates. 
	The specific design of the radiation baffles depends on the dimension and occupancy of the port. For example, the \IVC and \FCS ports, which must allow for a large gas flow, 
	have a series of double segmented baffles, while the \WT ports have \PTFE spirals that guide the wires while simultaneously blocking direct radition from the warmer stages. 

%% file: 4_cool_down.tex
\section{Cooldown systems}
\label{sec:cool_down}

	\begin{figure}[t]
		\centering
		\includegraphics[width=1.\columnwidth]{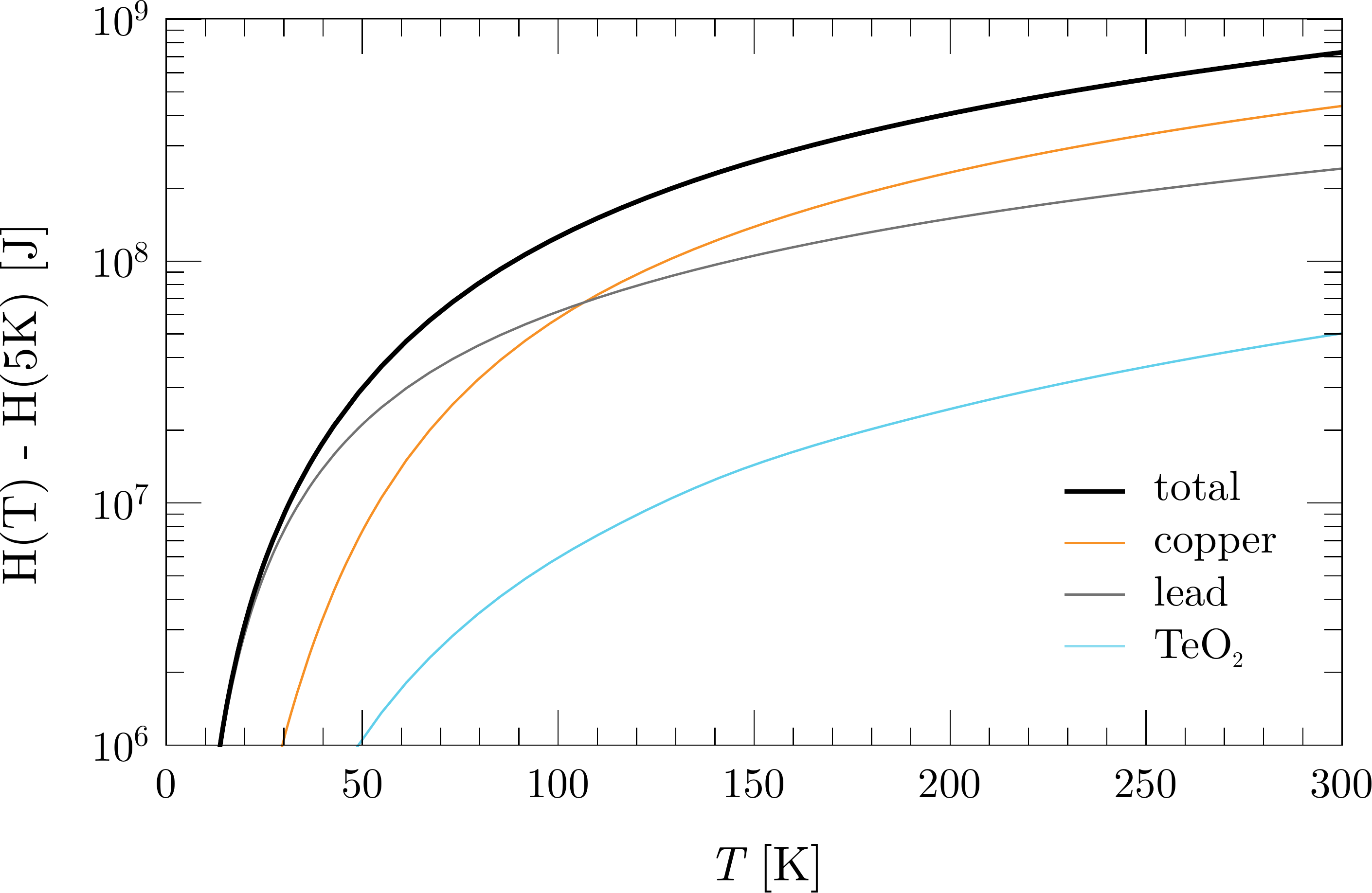}
		\caption{IVC enthalpy as a function of the temperature. 
			The contributions from the different materials are derived from the specific heat at different temperatures 
			of copper~\cite{Arblaster:2015Cu}, lead~\cite{Arblaster:2012Pb} and \ce{TeO_2}~\cite{White:1990} and combined with the masses 
			of each component (Table~\ref{tab:mass_inventory}).}
		\label{fig:IVC_enthalpy}
	\end{figure}

	The CUORE cryostat is cryogen-free and cooled by multiple PTs and a custom-built \ce{^3He}/\ce{^4He} dilution refrigerator,
	since ``standard'' solutions were not available for its size and mass.

	The {\PT}s provide enough cooling power to back the \DU and maintain the desired base temperature, but on their own would take too much time to bring the system down to 4\,K to 
	begin operation of the \DU. The total cooled mass is $13.7$\,t, of which $12.7$\,t is cooled to 4\,K or below (Tab.~\ref{tab:mass_inventory}). Between 300\,K and 40\,K, the cooling 
	system must remove $\sim 6.9\times10^8$\,J of enthalpy, which is over 95\% of the total enthalpy of the system, (Fig.~\ref{fig:IVC_enthalpy}.) The cooling power of each of 
	the five {\PT}s is $\gtrsim 100$\,W at 300\,K, but decreases quickly at lower temperatures; by 100\,K, the cooling power is already 50\,W. The {\PT}s alone, would imply 
	a cooldown time of order of a few months.
	
	Therefore, we designed a dedicated system to speed up the cooldown from room temperature, called the \FCS \cite{Pagliarone:2018mpj,Bucci:FCS_prep}. The system circulates \ce{He} 
	gas through a dedicated cooling circuit and injects the cold gas directly into the \IVC. The \FCS supports the {\PT}s during the initial stages of the cooldown, until all stages 
	inside the \IVC reach $\lesssim 100$\,K. The {\PT}s then continue the cooldown until the \DU can be activated and bring the system to base temperature.
	
	Examples of the full cooldown are shown later in Secs.~\ref{sec:Run4} and \ref{sec:CUORE_cool_down}.

\subsection{Fast cooling system}
\label{sec:FCS}

	\begin{figure}[tb]
		 \centering
		\includegraphics[width=.9\columnwidth]{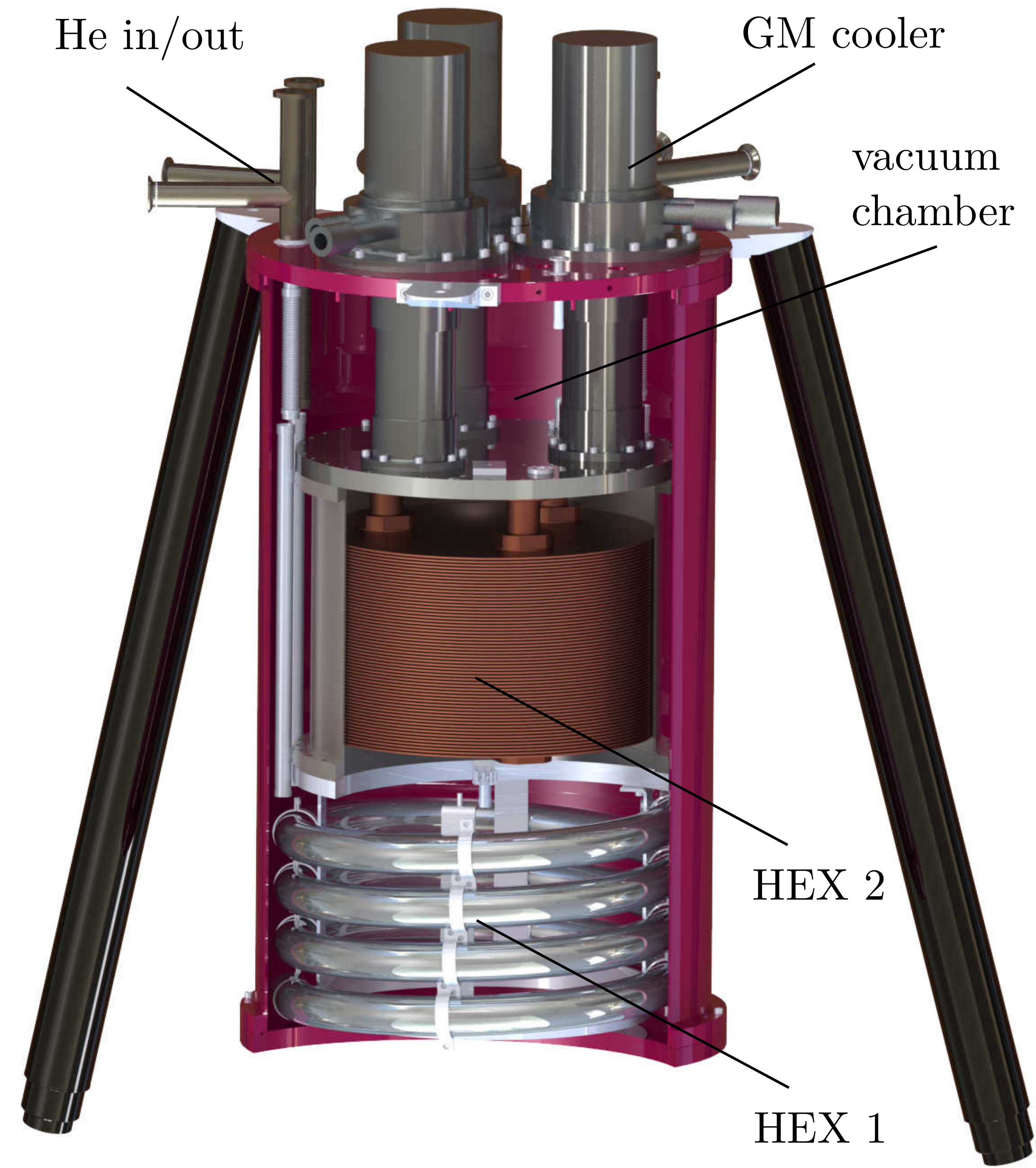}
		\caption{Rendering of the FCS cryostat.}
		\label{fig:FCS_cryostat}
	\end{figure}

	The \FCS consists of a separate cryostat cooled by three \GM cryorefrigerators AL600 made by Cryomech~\cite{Cryomech_AL600} (Fig.~\ref{fig:FCS_cryostat}), and a compressor that forces 
	the circulation of cold \ce{He} gas through a dedicated circuit and into the IVC.
	The GM cooling power is much higher than that of a PT, being $1800\,\W$ at $300\,\K$ and $350\,\W$ at $50\,\K$ and a single unit embedded in the CUORE cryostat would have sufficed 
	for the job. However, due to the lack of space and, especially, to the fact the GM would have created a significant thermal link between the 300\,K and the 4\,K stages during 
	the normal operation, this solution was not practical. Hence the need for an external cryostat and multiple units.

\begin{figure}[tb]
		\centering
		\includegraphics[width=.7\columnwidth]{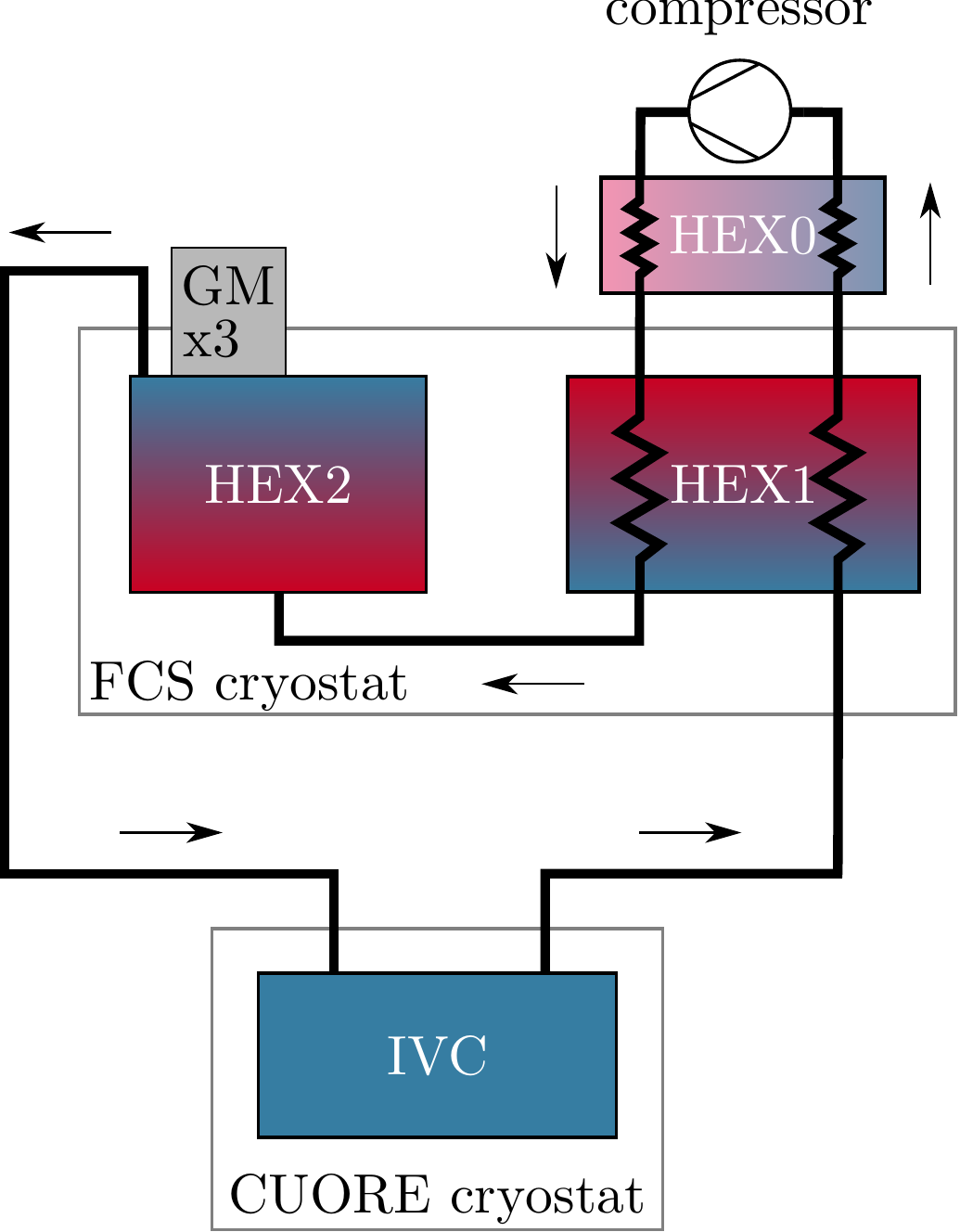}
		\caption{Schematic of the FCS gas circuit. The arrows indicate the direction of the \ce{He} flow.}
		\label{fig:FCS_circuit}
	\end{figure}

	A schematic of the circuit is shown in Fig.~\ref{fig:FCS_circuit}. The gas is cooled inside the \FCS cryostat inside two heat exchangers: a counter-flow heat exchanger (HEX~1), 
	and a series of Cu OFE plates directly connected to the cold stage of the {\GM}s (HEX~2). 
	In addition, a set of counter flow heat exchangers (HEX~0) precool the \ce{He} as it enters the \FCS cryostat.

	As the cold gas enters the \CUORE cryostat, it passes through an S-tube, which acts as a baffle to thermal radiation, and enters the \IVC. From the 4\,K plate, the cold 
	\ce{He} follows two paths. One guides the gas to the bottom of the \ILS, while the other guides the \ce{He} into the \HEX and \MC vessels through a set of \PTFE tubes 
	(chosen for low thermal conductivity and high radiopurity). The gas exits through a second port in the \IVC plate and returns to the \FCS cryostat.
	
	We placed important constraints on the \FCS:
	\begin{itemize}
		\item the \IVC pressure must never exceed $1.3\,\bbar$ to avoid any risk of damaging the cryostat vacuum seals;
		\item the circuit pressure must never be lower than the environmental pressure to prevent air from leak into the external pumping lines;
		\item the temperature inside the \FCS cryostat should always be lower than $202\,\K$ (the Radon melting point) while circulating -- 
			this creates a cryo-pump for \ce{Rn} and therefore prevents it from flowing into the \CUORE cryostat;
		\item excessive temperature gradients must be avoided across the detector, due to its intrinsic fragility, and across the IVC vessel, to prevent deformations of the vacuum vessel.
	\end{itemize}
	These constraints require significantly limiting the pumping speed of the compressor, and hence reducing the system cooling power.
	However, they ensure a safe cooldown for both the \CUORE cryostat and detector.
	Even in these conditions, the \FCS ensures a cooldown time within $\sim 20$ days (Figs.~\ref{fig:TopLead}, \ref{fig:R4_ILS} and \ref{fig:R5_TBase}).

	At the beginning of the cooldown, a compressor frequency of $\sim 20$\,Hz  provides a larger gas flow. As the temperature decreases, the compressor frequency has to be constantly 
	reduced down to a few Hz (Fig.~\ref{fig:Run4_FCS}).
	In the end, the minimum compressor frequency is limited by the parasitic power dissipation along the circuit, especially at the entrance/exit in the CUORE cryostat.
	Too low of a flow would result in a warm up of the circulating gas.
	Over the course of the cooldown, the temperature of the incoming \ce{He} goes from $\sim 150$\,K down to $\sim 50$\,K. In order to maintain a constant pressure while the 
	temperature decreases, we inject additional (clean) \ce{He} gas.
	
	The {\PT}s are turned on a few days after the FCS, typically when the \IVC temperatures are close to $200\,\K$. 
	By $\sim$50\,K, the {\PT}s dominate the total cooling power and the \FCS is turned off.
	The \ce{He} is pumped out of the \IVC , but we leave a few mBar of \ce{He} to act as an exchange gas between the 4\,K stage and the inner stages until they approach $\sim10$\,K.

\subsection{Pulse tubes}
\label{sec:PTs}

	\begin{figure}[t]
		\centering
		\includegraphics[width=1.\columnwidth]{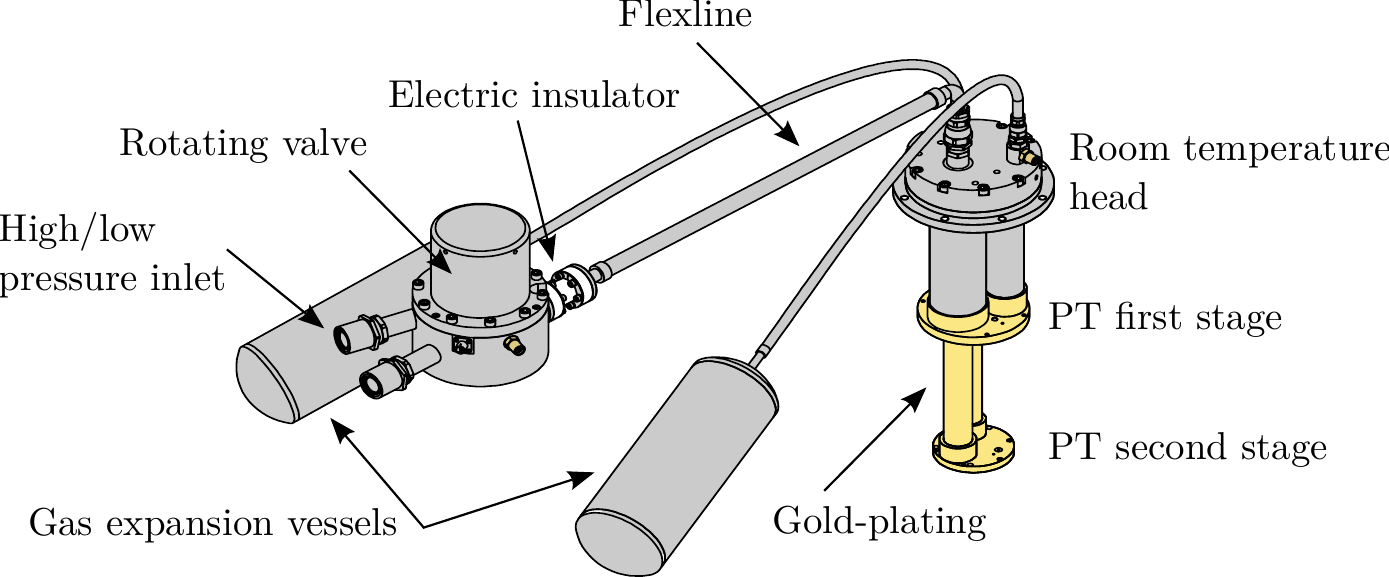}
		\caption{Schematic of a CUORE PT (PT415-RM by Cryomech).}
		\label{fig:PT}
	\end{figure}
	
	The \CUORE cryostat is cooled by five Cryomech PT415-RM pulse tube refrigerators~\cite{Cryomech_PT415} (Fig.~\ref{fig:PT}). The choice of a dry system relying on {\PT}s, 
	over the conventional ``wet'' design which relies on liquid helium (\ce{LHe}), was based on the both economic and logistic considerations. The wet system would require periodic \ce{LHe} refills, 
	which implies both a large cost, but also significant safety concerns given the amount of cryogen that would be needed. Instead, the dry system guarentees continuous operation, 
	thus increasing the total duty cycle of the experiment.
	
	The nominal cooling power of each PT is $32\,\W$ at $45\,\K$ (first stage) and $1.2\,\W$ at $4.2\,\K$ (second stage), while the lowest temperature achievable is close to $3\,\K$ 
	in case of no thermal load. The choice of the remote motor-head option and $2\,\text{ft}$ length of flexline between the \PT and motor-head, fundamental to reduce the 
	transmission of vibrations (see Sec.~\ref{sec:PT_dec}), implies a loss of $\sim 20$\% in cooling power for the \CUORE~{\PT}s.
	
	Given this cooling power, a minimum of three {\PT}s are required for normal operation. However, we decided to plan for the possibility that one {\PT} could fail over the course of 
	the experiment. This means accounting not only for the loss in cooling power, but also the additional heat load that an inactive \PT implies.%
	\footnote{The power dissipation for an inactive PT415 was computed by Cryomech to be $\sim 8.6$\,W at the first stage (between 300\,K and 40\,K) and $\sim 0.4$\,W at the 
		second stage (between 40\,K and 4\,K).}
	This required adding, not one, but two extra {\PT}s. During normal operation, one of the five {\PT}s is inactive.

	To maximize the conductance between the cooling elements and the cryostat, the \PT cold stages and the relevant sections of the cryostat plates are gold-plated. 
	The copper braids, which constitute the actual thermal link, have high thermal conductivity at low temperatures ($RRR \geq 100$) and are gold-plated as well.
	
	We can individually adjust the \ce{He} pressure driving each \PT, and optimize the cooling power of the system as a function of these pressures. Assuming working temperatures of 
	35\,K and 3.5\,K for the two thermal stages (see Sec.~\ref{sec:commissioning}), the cooling power of the 4 active {\PT}s is $\sim80$\,W and $\sim2$\,W, respectively.

\subsection{Dilution unit}
\label{sec:DU}
	
	\begin{figure}[tb]
		\hspace{100pt}
		\includegraphics[height=1.\columnwidth]{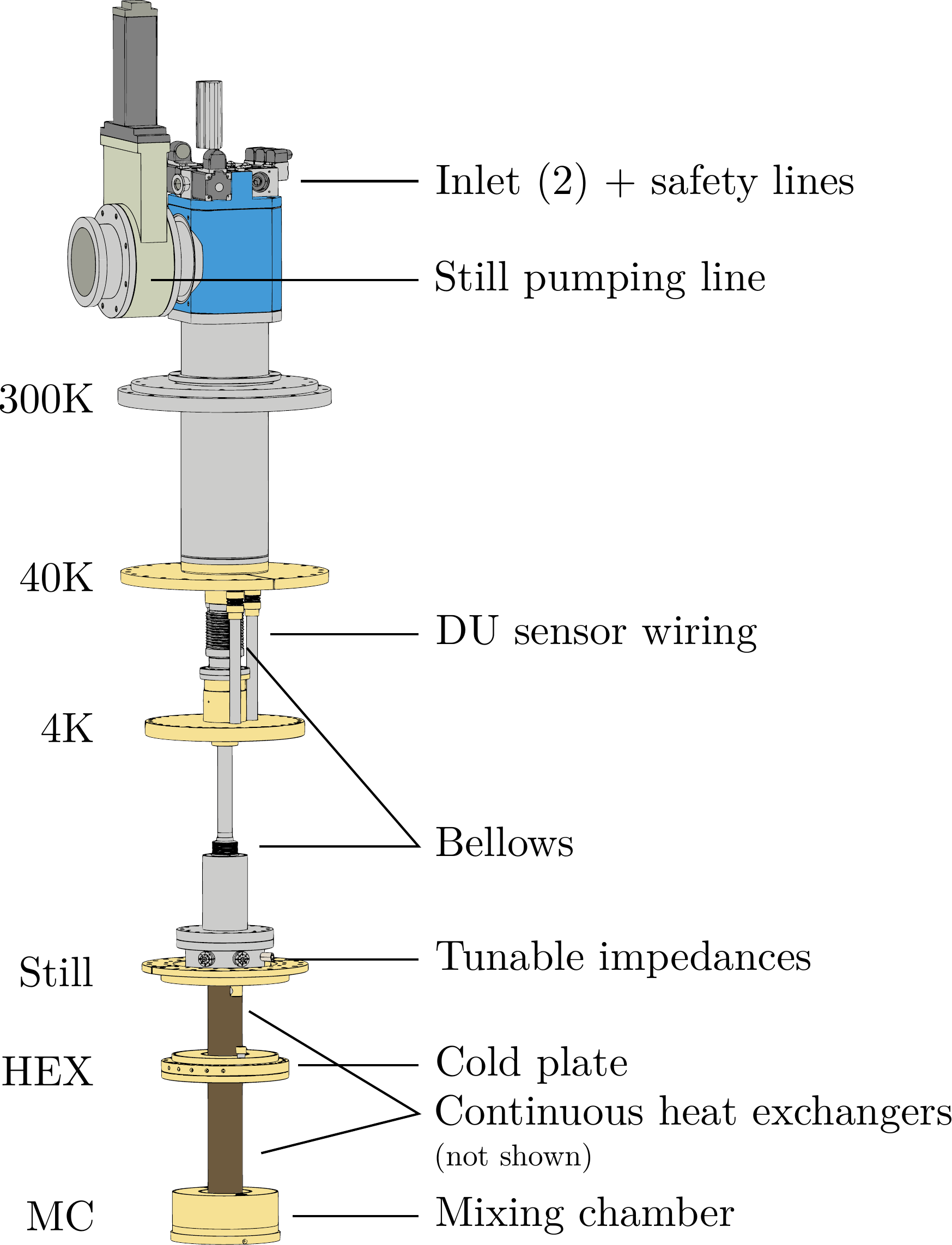} \\[-2pt]
		\caption{Schematic of the CUORE DU (custom CF-3000 by Leiden Cryogenics).}
		\label{fig:DU}
	\end{figure}

	The \CUORE \DU is a high-power custom CF-3000, manufactured by Leiden Cryogenics~\cite{LC_CF}. A schematic is shown in Fig.~\ref{fig:DU}. The DU is a standalone element that can 
	be inserted into the \CUORE cryostat or into a separate dedicated test cryostat. In both cases, each thermal stage of the \DU anchors to the corresponding plate. 
	Similar to the {\PT}s, the thermal connections are gold-plated to maximize the conductivity (the Still, HEX and MC plates are entirely gold-plated).
	
	The \CUORE \DU has two parallel condensing lines through which the \ce{^3He/^4He} mixture is injected. Each line is thermalized between the stages of a \PT, 
	and then passes through a thermalizer anchored to the 4\,K plate. 
	The thermalizer ensures that the incoming mixture is still cooled down to 4\,K even in the event that the PT that thermalizes the line fails.
	Below the 4\,K plate, the mixture passes through an impedance that cools it to $<1$\,K through Joule-Thomson expansion. The impedance can be tuned while the cryostat is open. 
	Apart from acting as a spare line for the mixture circulation during normal operation, the parallel condensing lines can be used simultaneously during the cooldown to 
	increase the mixture flow rate and achieve a higher cooling power. The maximum flow rate using both lines is $\gtrsim8\,\mmols$.

	The cooling power requirements on the \DU were set by the goal of operating stably at a temperature of $\sim10$\,mK.
	In a conservative approach, the target base temperature for the ``bare'' DU was set to $6\,\mK$.
	We extrapolated from simulations the cooling powers after the DU integration inside the \CUORE cryostat, starting from the estimate of the thermal loads on the different stages.
	The main contributions at the coldest stages comprise the detector wiring, the DCS (guide tubes + irradiation from the sources during calibration), the detector and lead 
	suspensions, and the cryostat support bars. We also added safety margins while computing the thermal loads, accounting for possible unknown sources besides the known ones.  
	
\begin{figure}[tb]
		\centering
		\includegraphics[width=1.\columnwidth]{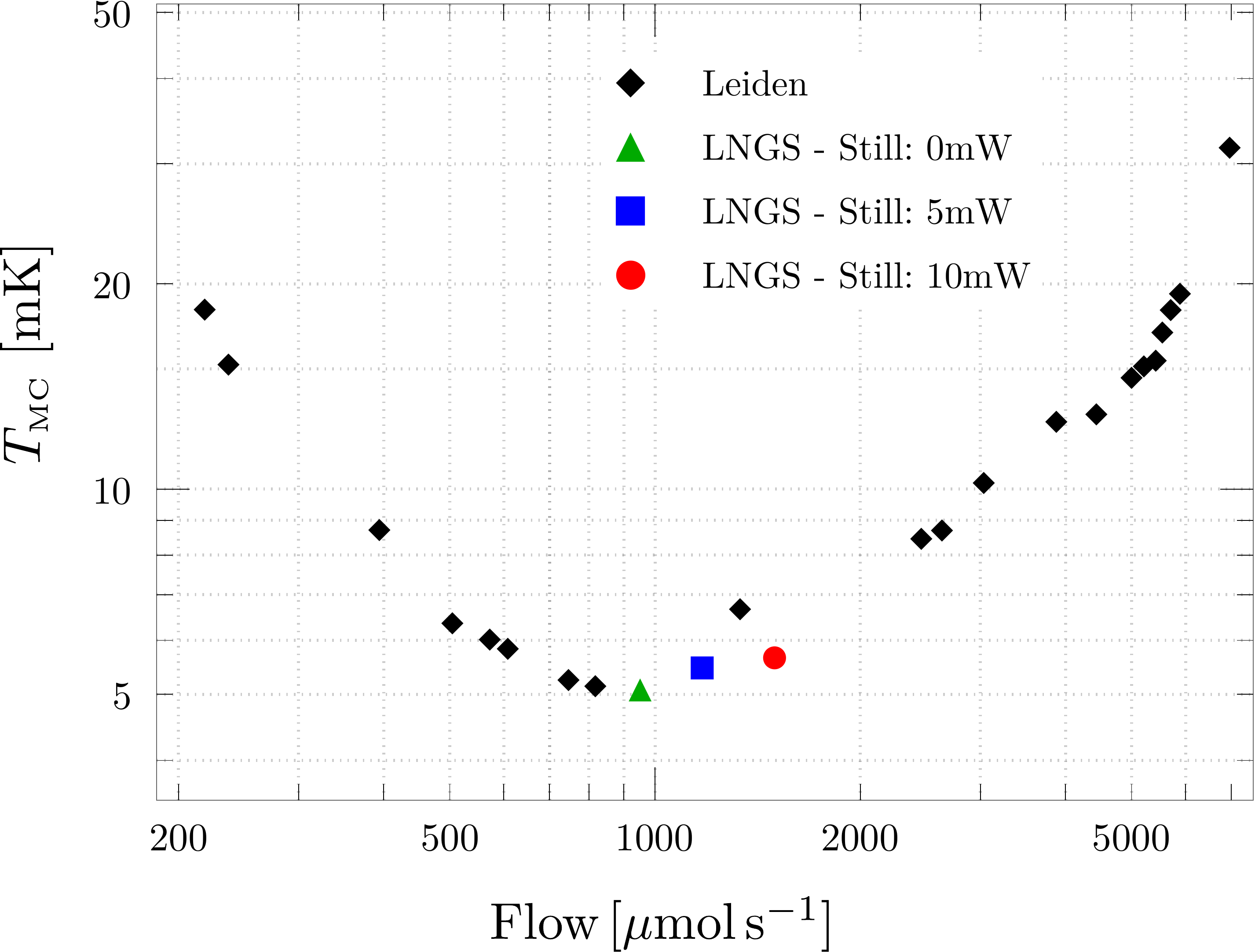}
		\caption[DU characterization ($T~vs.$ flow)]
			{Dependence of the MC temperature on the mixture flow. The black markers refer to the test performed at Leiden (August 2011). 
			The colored markers refer to the test performed at LNGS inside a test cryostat (April 2013) and are the same shown in the bottom panel of Fig.~\ref{fig:Q_vs_T} when no power 
			is injected on the \MC.}
		\label{fig:T_vs_F}
	\end{figure}

	The \DU was characterized in the test cryostat, both at the construction site (Leiden, The Netherlands) and at \LNGS. 
	The dependence of the base temperature on the mixture flow was determined by injecting different powers on the Still and measuring the MC temperature and flow rate. 
	The result is shown in Fig.~\ref{fig:T_vs_F} (black markers). The optimal value was found to be $\sim (800-1000)\,\mumols$.
	
\begin{figure}[tb]
		\centering
		\includegraphics[width=1.\columnwidth]{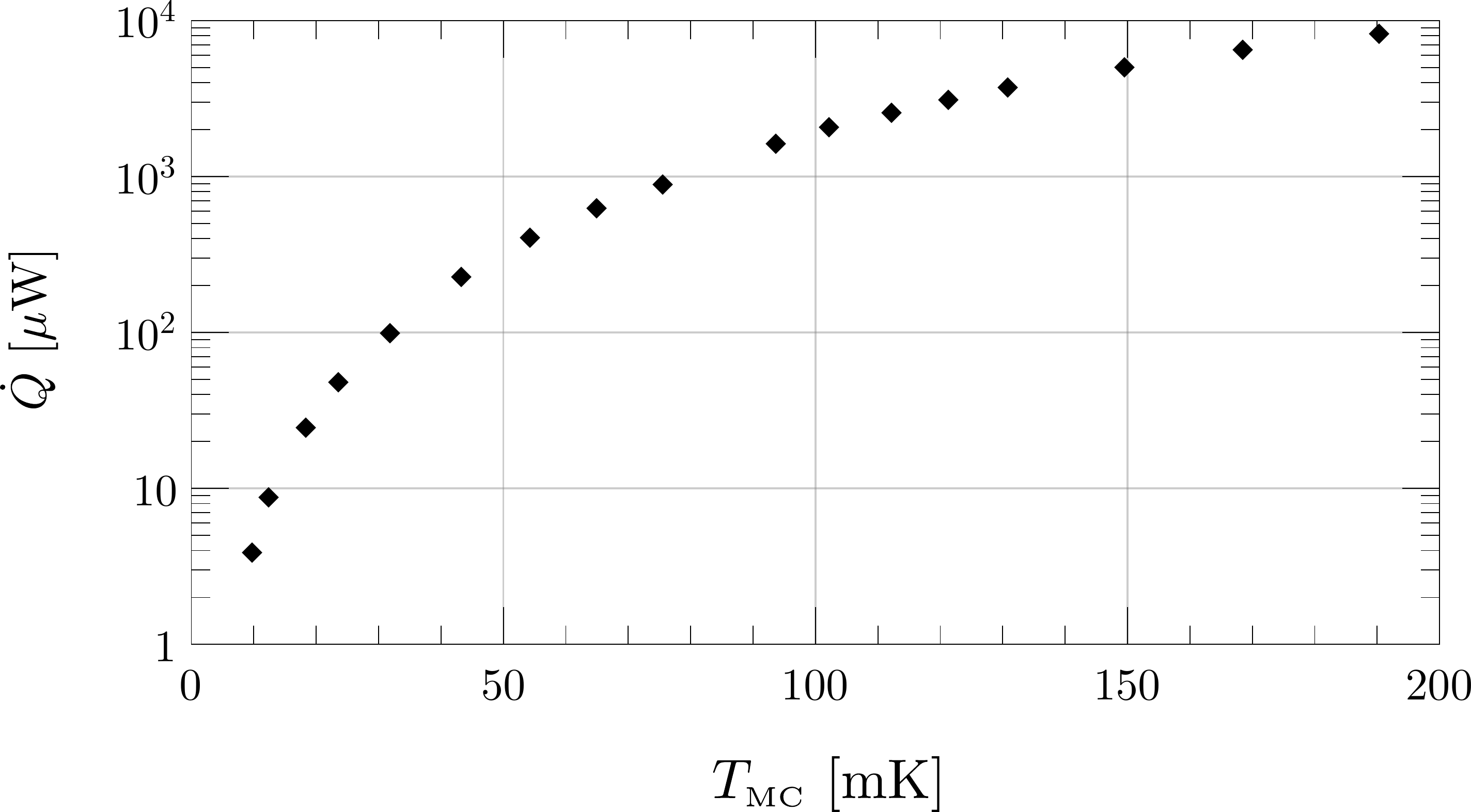}
		\\[+5pt] \phantom{ciao} \hspace{-1pt} 
		\includegraphics[width=.95\columnwidth]{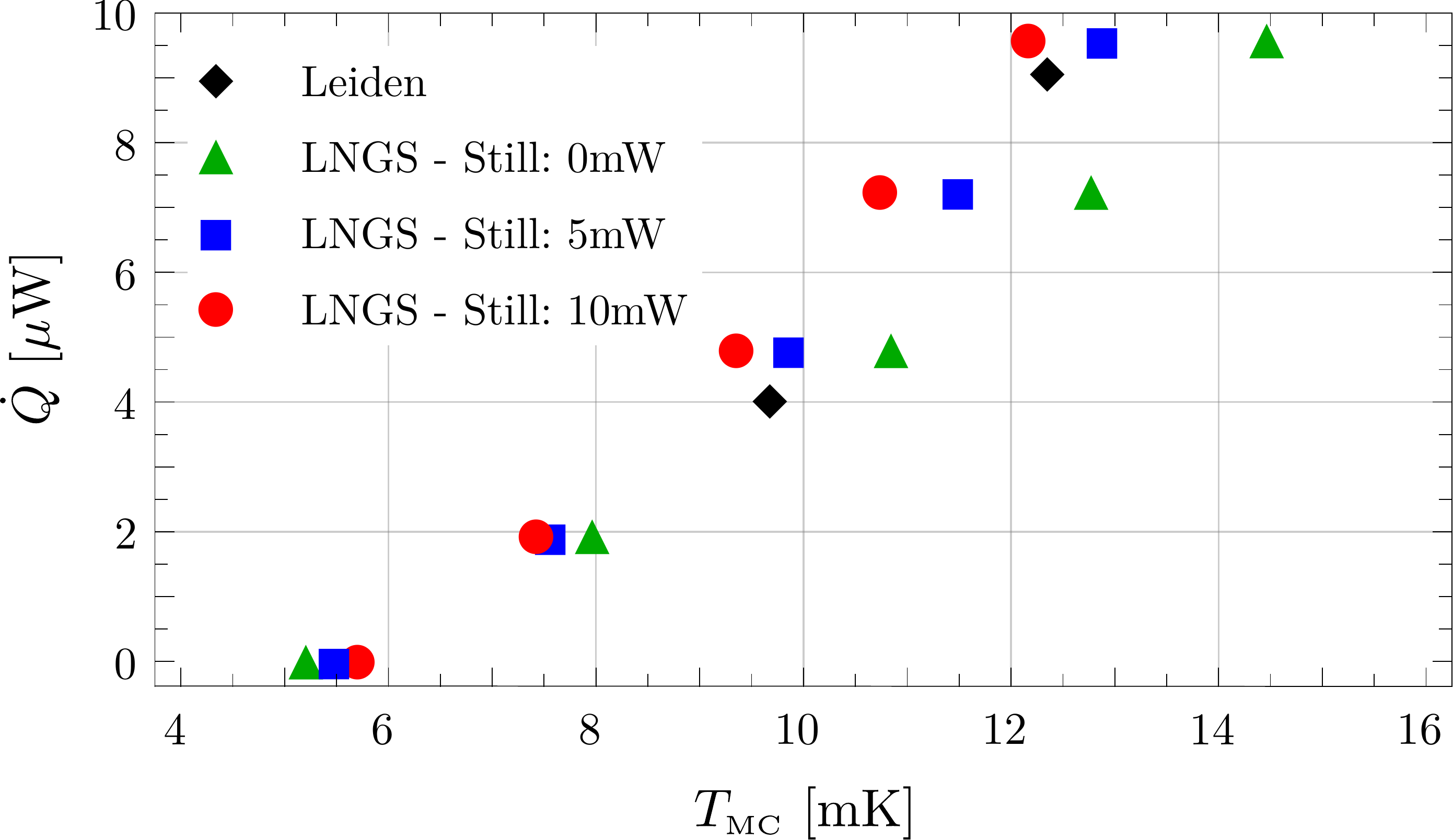}
		\caption{DU cooling power as a function of the MC temperature. (Top) Test performed at Leiden. 
			(Bottom) Zoom of the lowest temperature region with the inclusion of LNGS test results. The latter results show a better performance of the \DU for increased power on the still.}
		\label{fig:Q_vs_T}
	\end{figure}

	The cooling power of the \DU was measured by injecting power directly on the \MC stage (at fixed Still power) and observing the resulting base temperature. 
	The results of the Leiden characterization are shown in the top panel of Fig.~\ref{fig:Q_vs_T}.
	The obtained values were 2\,mW at 100\,mK, 3\,mW at 123\,mK and 4$\mu$W at 10\,mK -- these results actually exceeded the required specifications.
	The \LNGS characterization focused on the \DU behavior close to the base temperature, and the results are shown in Fig.~\ref{fig:T_vs_F}.
	The \LNGS results were compatible with the Leiden results when no power was injected on the Still, 
	while the \DU performance improved to $\sim 6\mu$W at 10\,mK with a mixture flow of $1500\,\mumols$.

%% file: 5_commissioning.tex
\section{Commissioning}
\label{sec:commissioning}

	\input{9_tab_commissioning}

	The cryostat commissioning proceeded in phases, with a series of runs characterized by specific sub-goals (Table~\ref{tab:cryostat_runs}).
	As new components were integrated into the cryostat (Table~\ref{tab:run_parts}), they were tested in a cooldown.

	In order to monitor the status during the cooldown and stable operation, more than $50$ thermometers have been installed in the CUORE cryostat for diagnostic purposes.
	The temperatures of the various components of the cryostat are monitored by multiple thermometers for redundancy.
	Given the broad range of values, from environmental temperature ($\sim 295$\,K) down to detector base temperature ($<10$\,mK), we use different sets of thermometers.
	Temperatures down to a few kelvin are recorded by commercial silicon diode resistance thermometers. For temperatures below 1\,K to a few tens of milikelvin, we use both commercial 
	and custom made ruthenium oxide resistor thermometers.
	Finally, on the \MC plate, we use an MFFT-1 Noise Thermometer by Magnicon~\cite{MFFT} as well as a Cerium Magnesium Nitrate thermometer to monitor temperatures all the way down to a few milikelvin. 
	These sensors are calibrated to a Superconducting Fixed Point device also installed on the \MC plate. 
	Below the Top Lead, bulk radioactivity places constraints on the sensor choice.
	Therefore, above a few kelvin, the detector temperature is monitored by ``bare chip'' diode thermometers.
	Once at base temperature, we use \NTD \ce{Ge} thermistors for both temperature monitoring and for temperature stabilization.

	In the end, the cryostat commissioning was performed over the course of 5 major cooldown runs, which required about four years to complete (see Ref.~\cite{Delloro_PhD-thesis:2017} for details).
	The commissioning ended with the successful installation and cooldown of the \CUORE detector. 

	\input{9_tab_cryostat_parts}

\subsection{Run 0}
\label{sec:Run0}

	The construction of the 4\,K outer cryostat, i.\,e.\ the 300\,K, 40\,K and 4\,K stages~\cite{Alessandria:2013ufa}, proceeded roughly in parallel with the separate commissioning of the \DU. 
	The goal of the first cooldown run, Run~0, was to verify the outer cryostat, both in terms of vacuum tightness and cooling power of the {\PT} system.

\begin{figure}[tb]
		\hspace{-20pt} 
		\includegraphics[width=1.07\columnwidth]{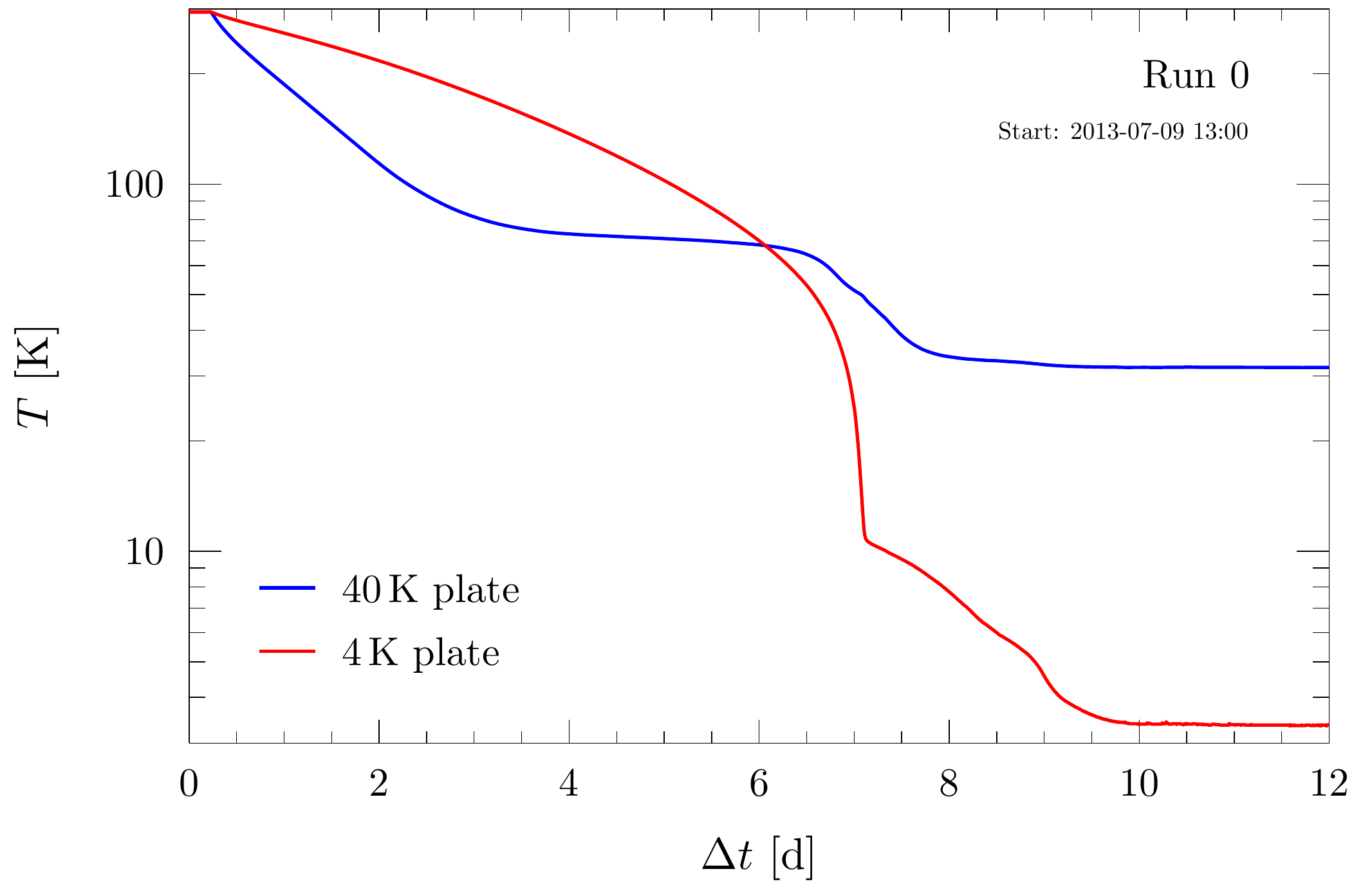}
		\caption{Temperatures of the 40\,K and 4\,K stages during the cooldown of the 4\,K outer cryostat in Run 0. The final temperatures of 32\,K and 3.4\,K are reached after 12 days. 
			Below $\sim 10$\,K (after day 7) the cooling rate is driven by a stainless steel mass that was being used to perform tests on the \DCS. The much higher heat capacity of the 
			stainless steel slows the cooldown relative to what it would have been in the absence of this test.}
		\label{fig:cool_down_Run0}
	\end{figure}

	The temperatures of the 40\,K and 4\,K plates from Run~0 are plotted in Fig.~\ref{fig:cool_down_Run0}. As it can be seen from the figure, the initial temperature drop is much steeper for 
	the 40\,K plate. This is due to the smaller mass at the 40\,K stage and to the higher cooling power of the PT first stage with respect to the second stage. 
	Then, the slope flattens some tens of kelvin below $\sim~100\,\K$. The cooling of the 4\,K proceeds more slowly and speeds up in the final part, after matching the 40\,K stage temperature.
		 
	During the cooldown, significant thermal gradients can form across the vessels. Temperature differences of up to $\sim40$\,K were measured between the bottom and top of the 40\,K vessel. 
	However, by the end of the cooldown, the temperature gradients were measured to be $\lesssim 0.2$\,K, which was in line with expectations from simulations.

	In Run~0, stable temperatures of 32\,K and 3.4\,K were reached on the 40\,K and 4\,K stages after about 12 days. Based on these temperatures and cooling time, 
	Run~0 was considered a successful test of the 4\,K outer cryostat.

\subsection{Run 1}
\label{sec:Run1}

	The Still, \HEX and \MC plates and vessels were installed in the next phase of commissioning, along with the integration of the \DU. 
	The Run~1 cooldown was a test of the performance of the ``complete'' cryostat.

\begin{figure}[t]
	\includegraphics[width=1.\columnwidth]{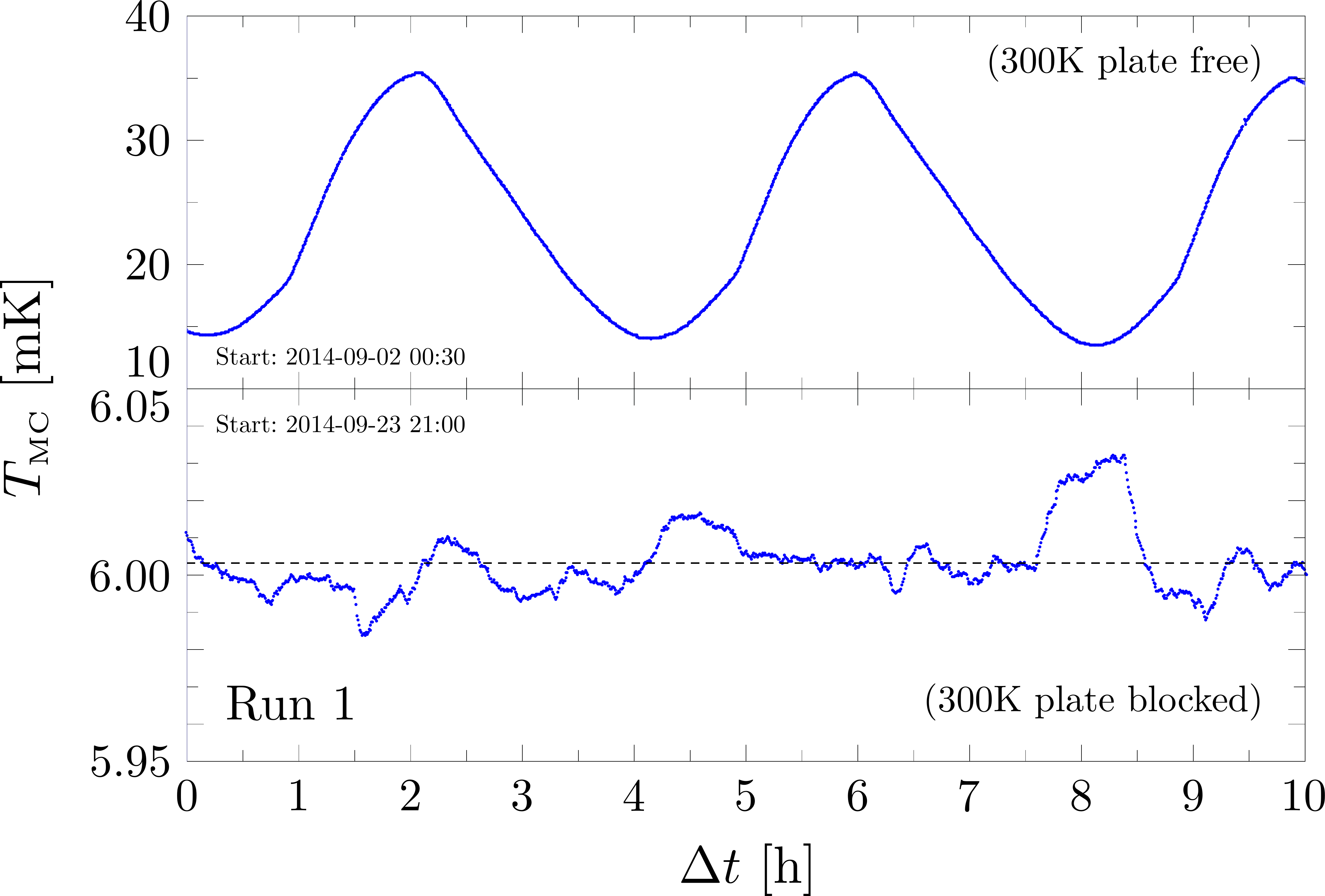}
	\caption{Temperature of the MC plate during Run 1. 
		(Top) The 300\,K plate has no constraints in the horizontal plane. the temperature oscillates over more than 20\,mK with a period of $\sim 4$\,h and does not reach 10\,mK. 
		(Bottom) The 300\,K plate is anchored to the MSP. The MC reaches a stable base temperature of $(6.00\pm0.01)\,\mK$.}
	\label{fig:TBase1_3}
	\end{figure}

	After $\sim 11$ days of cooldown, the 40\,K and 4\,K stages reached stable temperatures of 34.9\,K and 3.8\,K, at which point the \DU took over and completed the cooldown to base 
	temperature in less than one day.
	However, we initially observed large oscillations in the base temperature, reaching up to $\sim 35$\,mK with a $\sim 4$\,h period. This can be seen in the upper panel of Fig.~\ref{fig:TBase1_3}. 

	We determined that this instability was due to mechanical vibrations from the {\PT}s causing micro-friction and dissipating heat on the Still, \HEX and \MC stages. 
	This was tested through a series of diagnostic measurements with vibrational sensors installed on the 300\,K plate.

	To mitigate this effect, we installed a bracket system which rigidly anchored the 300\,K plate (see Sec.~\ref{sec:new_implementations}). 
	Three iron brackets were directly mounted to the 300\,K plate; each one pushing against another bracket, which was mounted to the \MSP. 
	This approach dramatically suppressed the temperature oscillations, as can be seen the bottom of Fig.~\ref{fig:TBase1_3}. 
	With these anchors installed, we were able to reach a satisfactory stable base temperature of 6\,mK, with the Still and \HEX at 1\,K and 35\,mK, respectively.

\subsection{Run 2}
\label{sec:Run2}

	The next step of the commissioning consisted of the installation of the \CUORE detector readout wiring, with the related boxes for the connection between the bolometers and 
	the front end electronics. This was tested in Run~2.
	
	The detector wiring consists of woven ribbon cables with twisted pairs of \ce{NbTi} wires in a NOMEX\sTM~texture, for a total of $\sim 2600$ wires~\cite{Andreotti:2009zza,Giachero:2013iya}.
	The wiring creates a link between the coldest stage and the external environment. Although the wiring had already been tested in dedicated setups, 
	it was fundamental to verify the actual impact on the cryogenic performance of the \CUORE cryostat.

	The wire thermalization down to 4\,K occurs via radiation transmission along the \PTFE spirals inside the \WT ports.
	Below, ribbons from the same \WT are sandwiched between gold-plated copper clamps installed above and below the Still and \HEX plates.
	On the \MC, the junction boxes for the connection between the ribbons and the cabling from the bolometers act as thermalizers.
	Measurements of the temperatures, both on the plates and along the wires, showed values in line with the expectations and the cooldown was considered successful.  

	Run~2 also provided the first possibility to run a bolometric detector inside the CUORE cryostat. The \MT~was the equivalent of a 2-floor CUORE tower and was mounted directly to 
	the \MC plate (rather than the \TSP, which was not yet installed). Since the \MC is coupled directly to the cryostat structure and is thus relatively noisy in terms of vibrations, 
	we did not expect to be able to observe physical events on the \MT. However, this was an opportunity to observed the behavior and performance of \NTD sensors.

	During the cooldown of Run~2, a superleak was discovered in the \DU circuit.%
	\footnote{A superleak is a leak that is permeable to superfluid \ce{^4He} due to its superfluidity, but not to \ce{^3He}.}
	This required extracting the \DU from the \CUORE cryostat and moving it back to its test cryostat, which can be thermally cycled much faster and allows for a much faster diagnosis. 
	After identifying the source of the leak, the \DU was repaired and reinstalled into the CUORE cryostat and we could move to the next step in the commissioning.

\subsection{Run 3}
\label{sec:Run3}

\begin{figure}[tb]
		\includegraphics[width=1.\columnwidth]{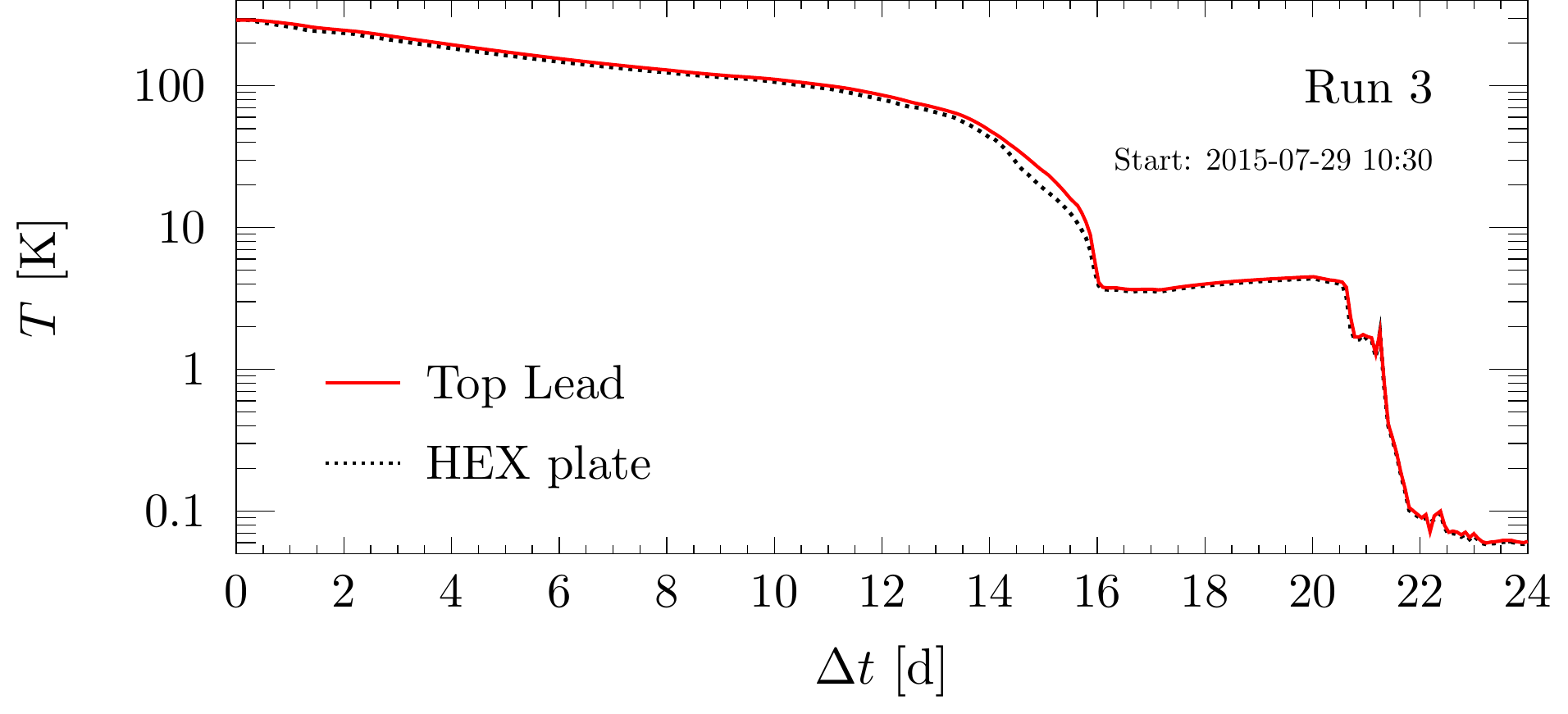}
		\caption{Temperature of the Top Lead during the cooldown of Run~3. The HEX plate temperature is shown as a reference. 
			The effect of turning off the FCS on day 14 is clearly visible as a slightly worsening of the lead thermalization process. 
			This immediately recovers once the lead becomes superconductive at $T_c\,\mbox{\footnotesize (Pb)} =7.2\,\K$.}
		\label{fig:TopLead}
	\end{figure}

	Run~3 tested the installation of the TSP and the Top Lead, which was assembled in an external cleanroom facility, and provided the first opportunity to employ the \FCS.
	The run proved successful. The cooldown time was consistent with previous runs and the addition of the new stages did not negatively impact the cryogenic performance. 
	The thermometry confirmed that the \TSP and the Top Lead were consistent with the stages to which they were thermalized. 
	The temperature of the Top Lead and \HEX stage (to which the Top Lead is thermalized) is plotted in Fig.~\ref{fig:TopLead}.

\subsection{Run 4}
\label{sec:Run4}

	\begin{figure}[tb]
		\includegraphics[width=1.\columnwidth]{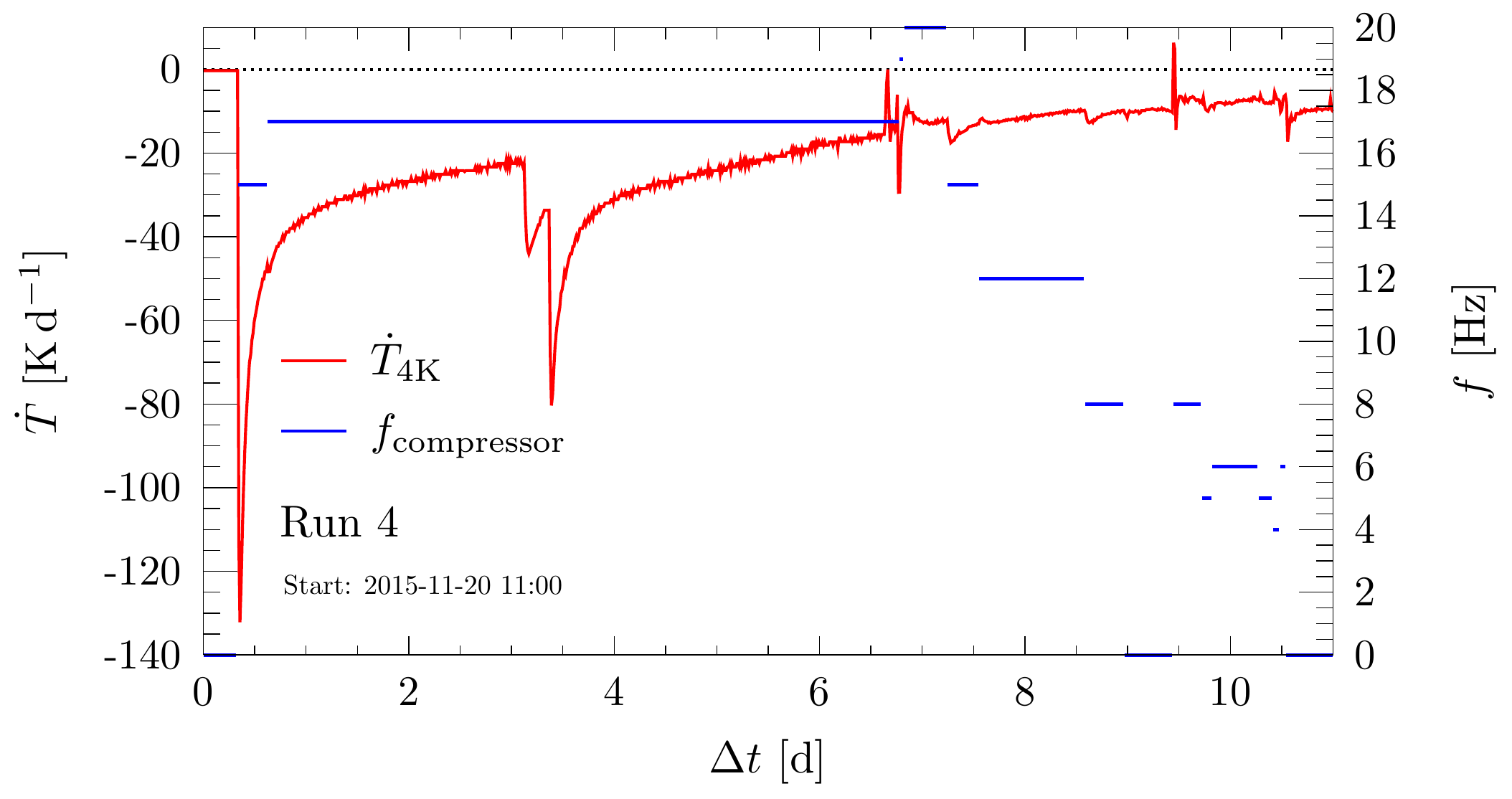} \\
		\phantom{ciao} \hspace{-28pt} 
		\includegraphics[width=.86\columnwidth]{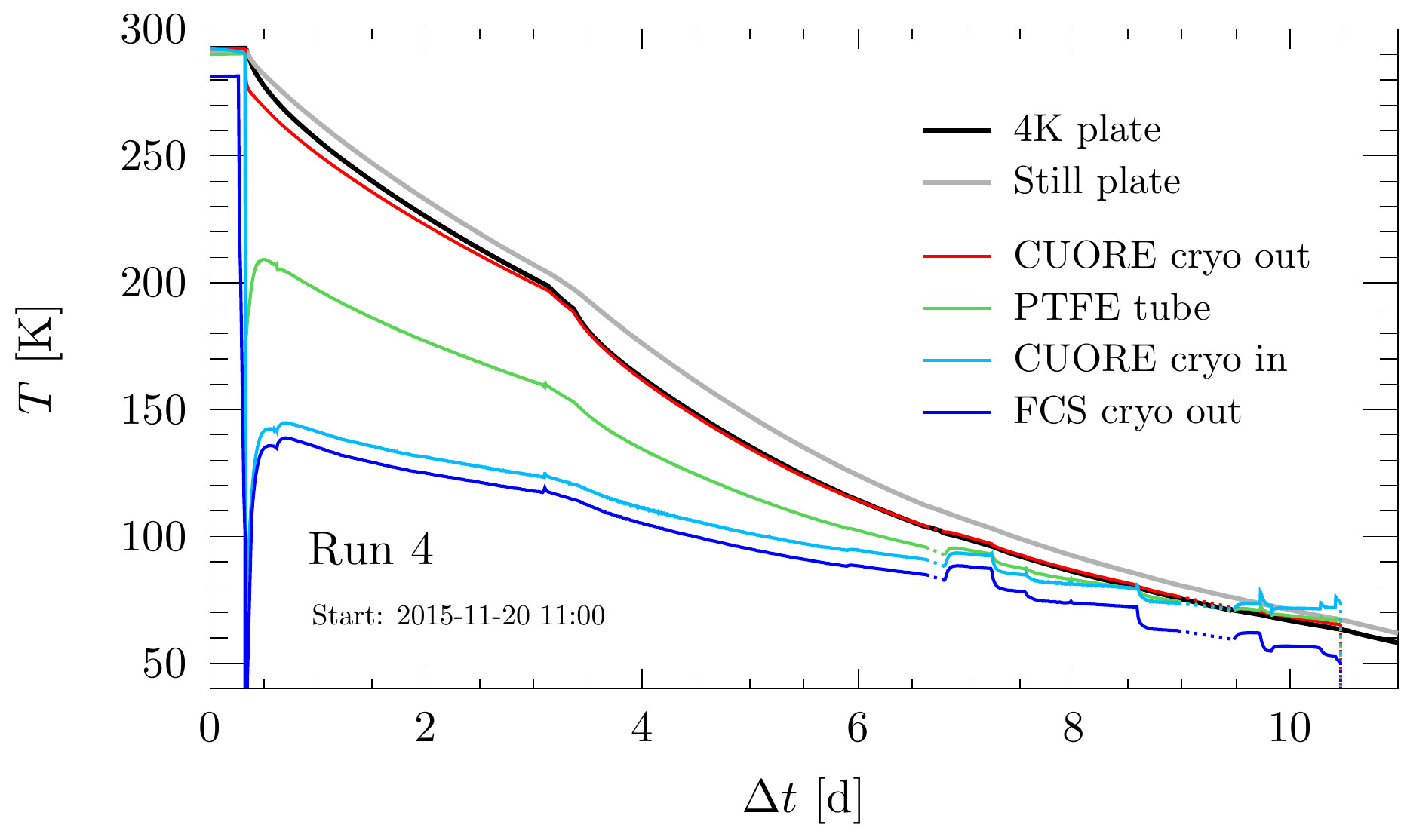} \\[10pt]
		\caption{(Top) Detail of the cooling rate with the FCS during Run 4. The compressor starts at a higher motor frequency, and this is periodically reduced over the course 
			of the \FCS operation. The improvement in the cooling obtained with the frequency tuning are visible on days 7, 8 and 9. 
			The PTs are turned on on day 3. The irregular behavior on days $6$ and $9$ is due to instabilities of the compressor cooling.
			(Bottom) Time profile of the \FCS temperatures in key points of the circuit. The 4\,K and Still temperatures are plotted as references.
			The \FCS cooling is effective until the temperature of the gas entering the \CUORE cryostat is no longer colder than the 4\,K/Still plates on day 9. 
			The difference between the temperatures measured at the exit of the \FCS cryostat and at the entrance of the \CUORE cryostat is caused by parasitic thermal loads on the 
			\FCS lines themselves.}
		\label{fig:Run4_FCS}
	\end{figure}
	
	The preparation for Run~4 began with the assembly of the \ILS~\cite{Bucci:ILS_prep}. After completion of Run~3 and the opening of the cryostat, the \ILS was installed. 
	This operation took place inside a dedicated cleanroom environment which had been assembled under the \CUORE cryostat.

	The cooldown of Run~4 was successful. The performance of the \FCS can be seen in  Fig.~\ref{fig:Run4_FCS}.
	The cooldown started with a higher compressor frequency, and therefore a larger gas flow. 
	The motor speed had to be reduced when the difference between the \ce{He} temperatures at the \IVC input and exit became too small.
	In fact, since the system was getting colder, the GM coolers were unable to keep up with the enthalpy being extracted from the cryostat.
	In the end,  the large thermal load of the cryostat, as well as thermal loads on the \FCS lines themselves, forced us to switch the \FCS off when the temperature of the 4\,K 
	stage dropped below the input \ce{He} temperature. 
	The {\PT}s were started on day 3 of the cooldown, which translated into a dramatic improvement in the cooldown speed on the cryostat. 
	We estimated that within one day, the {\PT}s were already providing $\sim$1/3 of the cooling power.
	
\begin{figure}[tb]
	\includegraphics[width=1.\columnwidth]{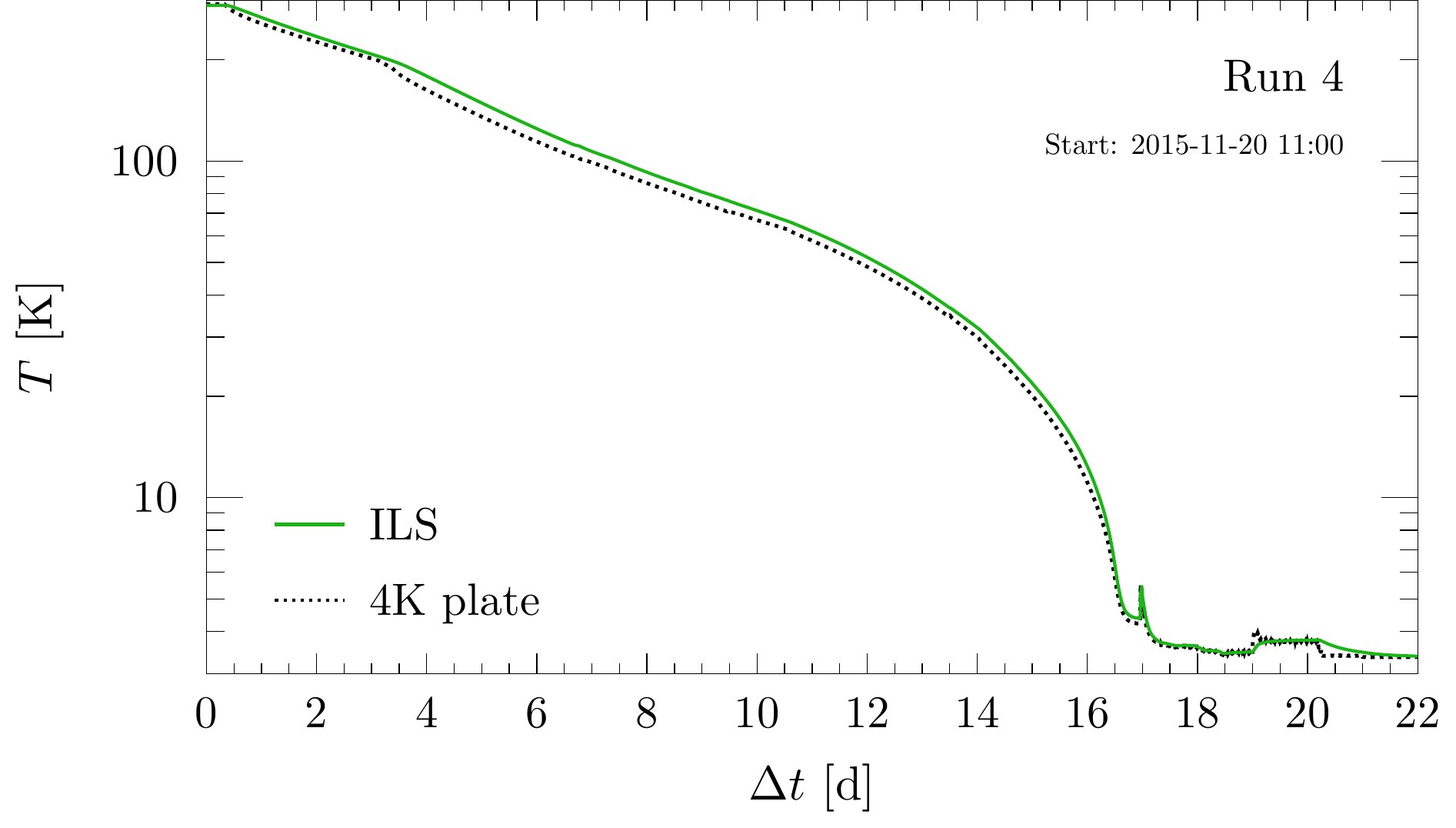}
	\caption{Temperature of the \ILS during the Run~4 cooldown. The 4\,K plate temperature is shown as a reference. 
		The effect of turning on the {\PT}s on day 3 is clearly visible as a faster cooling of the latter stage.}
	\label{fig:R4_ILS}
	\end{figure}

	Within about 22 days, the 40\,K and 4\,K stages reached stable temperatures of $34.5\,\K$ and $3.4\,\K$\, respectively. 
	The final temperature of the ILS was compatible with that measured on the 4\,K stage (Fig.~\ref{fig:R4_ILS}).\footnote{As can be seen in Fig.~\ref{fig:R4_ILS}, 
	we reached these temperatures within $\sim17$ days, however, due to some problems related to the compressor cooling water, it actually took a few more days to stabilize the temperature.}

	During Run 4, we performed a characterization of the \DU with the ``fully loaded'' cryostat (with the exception of the detector). 
	The new measured cooling power was close to that originally measured (Fig.~\ref{fig:Q_vs_T}), despite the additional thermal loads. 
	Its value was $3\,\upmu$W at $10$\,mK, obtained with an optimal mixture flow within the range $(1010-1020)\,\mumols$.
	The long duration of Run~4 also allowed us to test the stability of the \MC base temperature on the timescale of a few months, which maintained an average value of 6.3\,mK.
	
	In Run~4, the \MT~was mounted to the \TSP, identically to the way the eventual \CUORE detector would be. 
	A large part of this run was devoted to the understanding of the noise generation and propagation through real-time analysis of the noise on the \MT~bolometers. 
	By monitoring the variation in intensity and frequency of noise peaks in the bolometer noise power spectra, it was possible to identify sources of vibrational and mechanical 
	noise and find solutions to mitigate them. Many of these solutions pushed in the direction of increasing the rigidity of the cryostat structure.

\subsection{New implementations after the commissioning}
\label{sec:new_implementations} 

	As discussed in Sec.~\ref{sec:cryo_design}, the design of the \CUORE cryostat had to satisfy a series of demanding and sometimes countervailing requirements. 
	The outcome was a cryogenic infrastructure that is unique in its kind. In this sense, the \CUORE cryostat is a prototype. From a practical point of view, this means that it 
	was not always possible to accurately foresee every challenge at the design stage. To this extent, the commissioning phase proved invaluable in terms of gaining the experience to diagnose, 
	correct and operate the \CUORE cryostat. It also allowed us to modify some choices made in the initial design.
	
	The major adjustment with respect to the initial cryostat design concerned the rigidity of the support structure. The suspension system was designed to isolate the cryostat 
	infrastructure from mechanical vibrations, however, during the commissioning runs, this proved to present a problem for maintaining a stable base temperature. We determined that 
	the entire cryostat was behaving like an articulated pendulum, and the friction generated by the relative movement of plates was creating a large heat load, 
	leading to an unstable base temperature. These oscillations were being driven by the {\PT}s, despite the attempts to suspend the \PT motor heads and dissipate these forces elsewhere.
	
	Among the modifications to counter these oscillations, we decided to prevent relative motion of the plates by fixing parts of the joints, and to reduce the motion of the 300\,K 
	plate by fixing it to the \MSP (Sec.~\ref{sec:Run1}). These solutions provided a more efficient path to dissipate the energy imparted by the {\PT}s rather than allowing it to grow into 
	a sizable oscillation. This was necessary to operate a bolometric detector, as became evident during the operation of the \MT.
	
	We also observed that the seismic dampers supporting the entire cryostat infrastructure (see Fig.~\ref{fig:support_structure}), had resonant oscillation frequencies in the $(1-3)$\,Hz 
	range, which lies right in the middle of the \CUORE signal bandwidth. To mitigate this, we designed and installed a ``mechanical fuse'' system, which prevents the growth of 
	small oscillations, but will break away in the case of a strong acceleration (such as an earthquake). 
	This consists of a series of thin steel plates inserted between the concrete support structure and the foundation. 
	This system allows us to reduce mechanical vibrations, while still satisfying the seismic safety of the structure.
	
	After these changes, we could finally proceed towards the installation of the \CUORE detector.

%% file: 9_tab_commissioning.tex
	\begin{table}[t]
	\caption[CUORE cryostat commissioning at LNGS (table)]
		{Summary of the CUORE cryostat commissioning runs at LNGS. For each test run, goals and schedule are reported.}
		\begin{center}
		\small{
		\begin{ruledtabular}
		\begin{tabular}{l l l}
		Run	&Duration						&Goals										\\[+1pt]
		\hline\\[-10pt]
				&\Oct 2012 - \Jan 2014		&- DU standalone tests					\\[+7pt]

		0		&\Jul 2012 - \Aug 2013		&- reach stable PT temperature		\\[+7pt]
		
		1		&\Sep 2013 - \Oct 2014		&- merge 4\,K cryostat and DU			\\
				&									&- installation of inner plates		\\
				&									&\hdot~and vessels						\\
				&									&- reach stable base temperature		\\[+7pt]

		2		&\Oct 2014 - \Jan 2015		&- installation of detector read-out\\ 
				&									&\hdot~wiring						 		\\[+7pt]

		3		&\Jan 2015 - \Sep 2015		&- Top Lead installation				\\
				&									&- TSP installation						\\
				&									&- cooldown with FCS						\\[+7pt]

		4		&\Jul 2015 - \Mar 2016		&- ILS	installation					\\
				&									&- \MT~performance	 					\\
				&									&\hdot~optimization						\\[+1pt]

		\end{tabular}
		\end{ruledtabular}
			}
		\end{center}
	\label{tab:cryostat_runs}
	\end{table}

%% file: 9_tab_cryostat_parts.tex
	\begin{table}[ptb]
		\centering
		\caption[CUORE cryostat part installation (table)]
			{CUORE cryostat part installation during the commissioning runs. The symbols \yes/\no~indicate
			presence/absence of the various elements.}
		\begin{ruledtabular}
		\begin{tabular}{l rrrrr}
			Part								&Run 0			&Run 1				&Run 2			&Run 3			&Run 4	\\
		\hline
		\\[-7pt]
		\emph{Load structure} \\		\cline{1-1} \\[-7pt]

			300\,K, 40\,K, 4\,K			&\yes				&\yes					&\yes				&\yes				&\yes		\\
			Still, HEX, MC		 			&\no				&\yes					&\yes				&\yes				&\yes		\\
			TSP			 					&\no				&\no					&\no				&\yes				&\yes		\\

		\\[-7pt]
		\emph{Lead shields} \\		\cline{1-1} \\[-7pt]
			Top Lead 						&\no				&\no					&\no				&\yes				&\yes		\\
			ILS		 						&\no				&\no					&\no				&\no				&\yes		\\

		\\[-7pt]
		\emph{Cooldown} \\			\cline{1-1} \\[-7pt]
			DU				 					&\no				&\yes					&\yes				&\yes				&\yes		\\
			PTs								&3					&4						&4					&5					&5			\\
			FCS 								&\no				&\no					&\no				&\yes				&\yes		\\
		\\[-7pt]
		\emph{Read out} \\			\cline{1-1} \\[-7pt]
			WT ports							&\yes				&\yes					&\yes				&\yes				&\yes		\\
			detector wiring				&\no				&\no					&\yes				&\yes				&\yes		\\
			\MT								&\no				&\no					&\yes				&\yes				&\yes		\\

		\end{tabular}
		\end{ruledtabular}
	\label{tab:run_parts}
	\end{table}

%% file: 6_CUORE_cool_down.tex
\section{CUORE cooldown}
\label{sec:CUORE_cool_down}

	\begin{figure}[t]
		\includegraphics[width=1.\columnwidth]{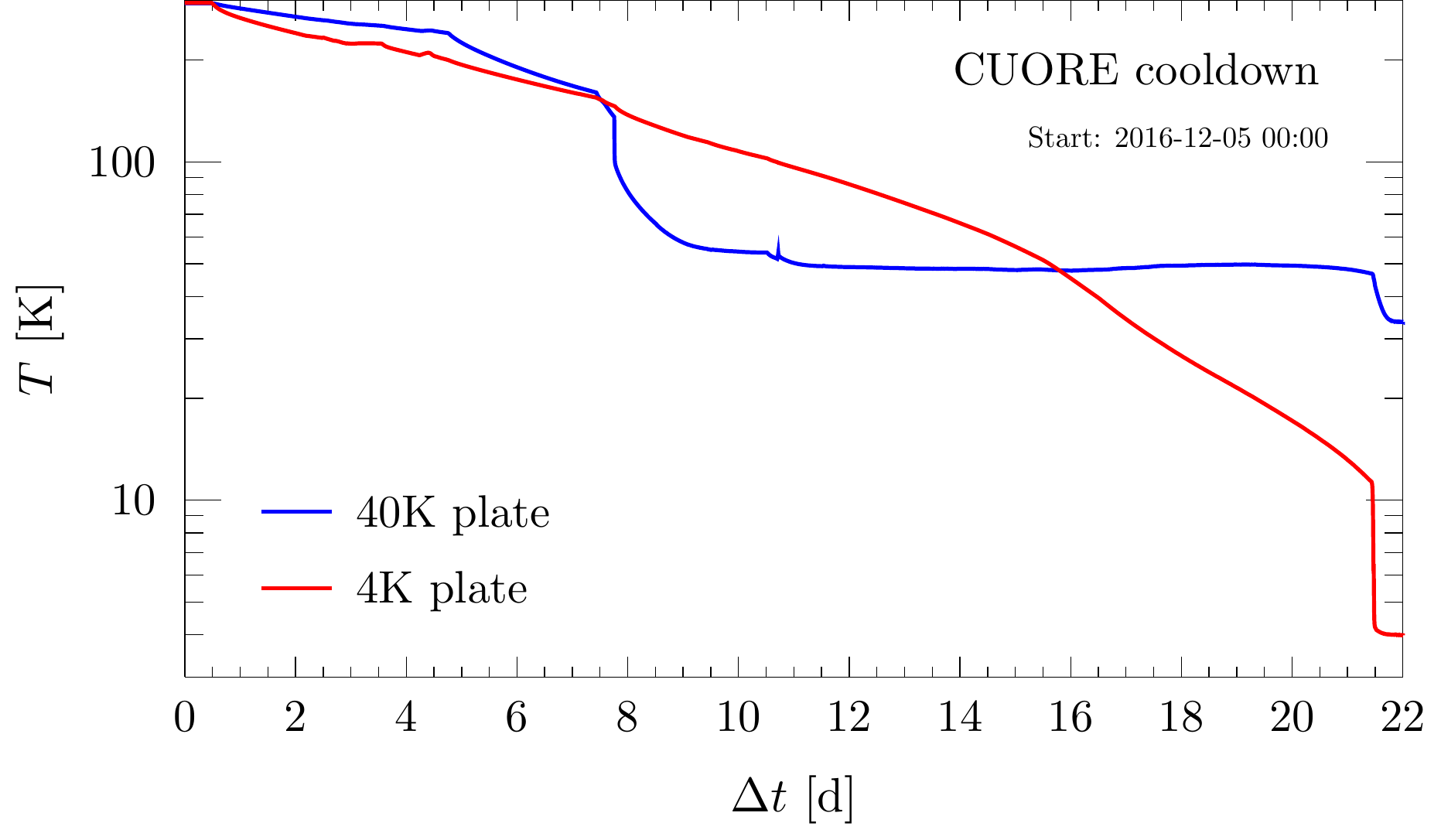}
		\\ \phantom{ciao} \hspace{0pt} 
		\includegraphics[width=1.\columnwidth]{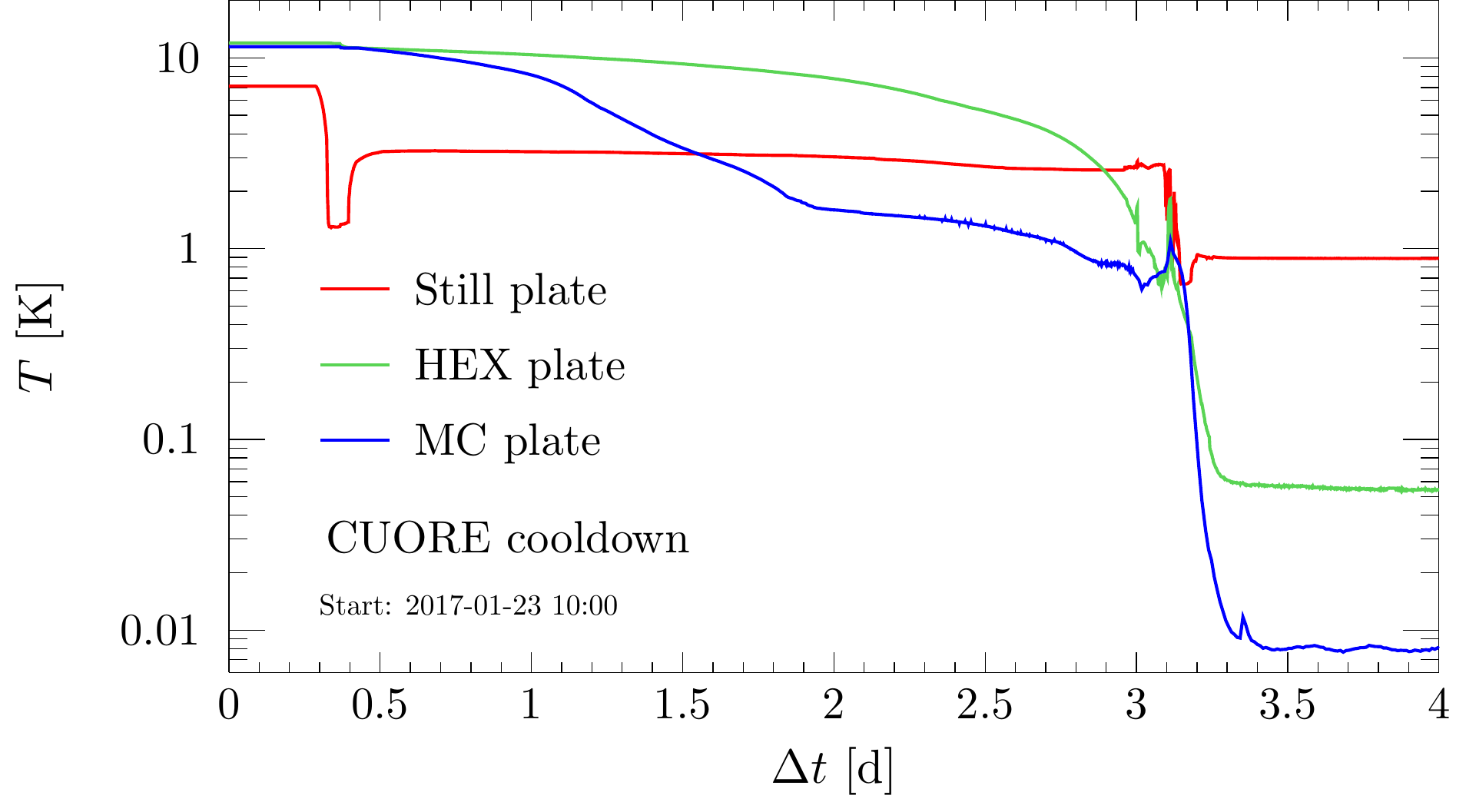}
		\caption{CUORE cooldown. (Top) Temperatures of the 40\,K and 4\,K stages as a function of time since the turn on of the FCS. 
			The stages reach temperatures of 33.6\,K and 4.0\,K after $\sim 22$\,d.
			The sudden changes in cooling rates of both stages seen around day 4 is due to the starting of the PTs. The \FCS was turned off on day 10.
			(Bottom) IVC stage temperatures as a function of time since the start of the DU circulation.
			The base temperature achieved on the MC with the full experiment apparatus is 8\,mK, with 0.89\,K and 55\,mK on the Still and HEX stages, respectively. 
			The initial Still temperature drop and rise within the first day is due to the switch from \ce{^4He} to \ce{^3He} circulation. 
			By day 3, the gas has fully condensed and the dilution cooling takes over. This is visible as the small spike in the MC temperature.}
		\label{fig:R5_cool_down}
	\end{figure}

	After the installation of the \CUORE detector, the closing of the cryostat was made significantly more complicated. The mounting of the towers to the \TSP took place 
	inside a dedicated cleanroom environment, which was flushed with Rn-free air~\cite{Benato:2017kdf}. This cleanroom was kept in place while closing the inner cryostat vessels up 
	to the 4\,K shield. Once the 4\,K vessel was closed, the \IVC could be pumped and the detector could be protected from contact with \ce{Rn} by the vacuum.
	Once closed, both the 4\,K and 40\,K vessels and plates had to be covered with new layers of superinsulation.
	The old superinsulation that had been in use for the cryostat commissioning, and was now dirtier after years of exposure to air and continuous handling,
	had been removed to allow the cleaning of the vessels.

	The progression of the \CUORE cooldown is shown	in Fig.~\ref{fig:R5_cool_down}. 
	As usually, this began with the \FCS, while the {\PT}s were turned on after a few days. 
	The \FCS was then turned off and disconnected once the thermal stages reached $\sim$150\,K.
	As expected, the cooldown took $\sim$ 20 days before turning on the \DU unit.
	We paused the cooldown for about one month to perform some debugging of the detector electronics and to address a leak that had formed in the dilution circuit (outside the cryostat). 
	Once the \DU was turned on, it took less than 4 days to reach the base temperature of 8\,mK on the \MC, with 0.89\,K and 55\,mK on the Still and \HEX stages respectively.
	
\begin{figure}[tb]
		\includegraphics[width=1.\columnwidth]{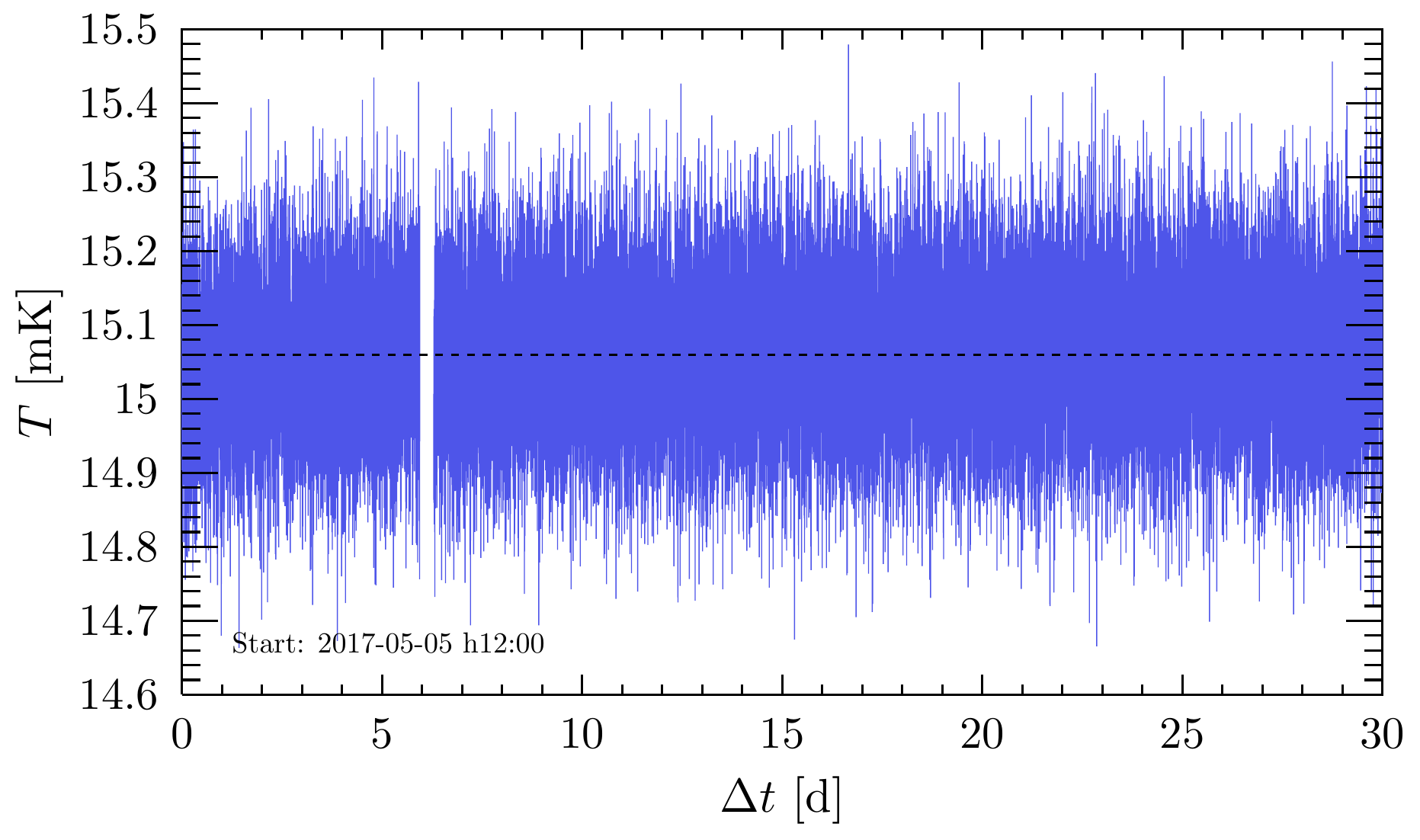}
		\caption{Base temperature stability over a thirty day period while the cryostat was ``undisturbed''  during the initial phase of the CUORE data taking. 
			The MC temperature is $(15.06\pm 0.10)\,\mK$. The dashed line indicates the average value. The gap on day 6 is due to hardware operation the cryogenic system.}
		\label{fig:R5_TBase}
	\end{figure}

	An initial scan of the detector behavior at various temperatures identified 15\,mK as a suitable operating temperature.
	Fig.~\ref{fig:R5_TBase} shows the stability of the system for over one month of operation.
	The \CUORE cooldown and the beginning of the detector operation successfully concluded the commissioning of the \CUORE cryostat.
		
	The analysis of the first collected physics data provided a measurement of the background in the \NDBD decay region~\cite{Alduino:2017ehq}.
	The observed value, $(0.014 \pm 0.002)$\,counts\,keV$^{-1}$\,kg$^{-1}$yr$^{-1}$, was in line with the expectations derived from the background model~\cite{Alduino:2017qet}.
	The strict material selection and the dedicated construction techniques for the cryostat components resulted in a cryogenic system fully compliant with the design goals.

%% file: 7_summary.tex
\section{Summary}
\label{sec:summary}

	The \CUORE cryostat was designed to satisfy a wide set of very stringent requirements, ranging from cryogenic performance and duty cycle, to vibration isolation and 
	mechanical stability, and to low radioactive content and shielding. 
	The full construction and assembly of the cryostat itself took more than three years. 
	The commissioning phase was long and complex, requiring several cooldowns to integrate the numerous custom components.

	On the cryogenics side, the success of the \CUORE cryostat was demonstrated by the cooldown of the \CUORE detector to 8\,mK, and by the ability to operate stably at 15\,mK.
	At the same time, the measured background in the \NDBD decay region, $(0.014 \pm 0.002)$\,counts\,keV$^{-1}$\,kg$^{-1}$yr$^{-1}$, is in line with the \CUORE sensitivity goal
	and does not show any excess due to the cryogenic system.
	This proves the effectiveness of the material selection and construction techniques for the cryostat components.

	The \CUORE cryostat has set the stage for one of the most sensitive searches for \NDBD. More generally, it demonstrated the ability to cool tonne-scale detectors 
	to temperatures of a few mK, and operate them in low-noise and low-background environments compatible with the requirements for rare-event searches.
	This marks a major milestone in the history of low-temperature detector techniques.

%% file: 8_Acknowledgments.tex
	\begin{acknowledgments}
		We wish to express our gratitude to the entire CUORE Collaboration for the invaluable support.
		Thanks are due to the Laboratori Nazionali del Gran Sasso for generous hospitality.
		We thank Leiden Cryogenics for designing and constructing our high-performance Dilution Unit and the Fast Cooling System, 
		and the patient and continuous support during the cryogenic system commissioning.
		We gladly acknowledge the help of S.\ Parmeggiano, P.\ Mereu and G.\ Giraudo in various stages of the construction.
		We thank Low T-Solutions and TCService s.\,p.\,a.\ for the ingenuous and continuous support in the commissioning phase.
		We would like to thank A.\ Benoit, M.\ B\"uhler, P.\ Camus, E.\ Coccia, C.\ Gargiulo, P.\ van der Linden, N.\ Mirabolfathi, P.\ Negri, P.\ Rapagnani and A.\ Woodcraft 
		for the reviews and invaluable feedback on the design and implementation of the cryogenic systems, and the members of the LNGS Scientific 
		Committee and of the international CUORE Review Committee for the continuous oversight and support during the design, construction, and operations of the CUORE experiment.
		This work was supported by the Istituto Nazionale di Fisica Nucleare (INFN); the National Science Foundation under Grant Nos. NSF-PHY-0605119, NSF-PHY-0500337, NSF-PHY-0855314,
		NSF-PHY-0902171, NSF-PHY-0969852, NSF-PHY-1307204, NSF-PHY-1314881, NSF-PHY-1401832, and NSF-PHY-1404205; the Alfred P. Sloan Foundation; the University of Wisconsin Foundation; 
		and Yale University. This material is also based upon work supported by the US Department of Energy (DOE) Office of Science under Contract Nos. DE-AC02-05CH11231, DE-AC52-07NA27344, 
		and DE-SC0012654; and by the DOE Office of Science, Office of Nuclear Physics under Contract Nos. DE-FG02-08ER41551 and DE-FG03-00ER41138.  This research used resources of the National
		Energy Research Scientific Computing Center (NERSC).
	\end{acknowledgments}

%% file: 9_glossary.tex
\section{Glossary}
\label{sec:intro}

	Here, we list the acronyms related to the CUORE cryogenic system and its components that have been introduced and used throughout this work:

	\begin{flushleft}
		\begin{tabular}{l l}
		
	DCS		&Detector Calibration System 	\\
	DU			&Dilution Unit						\\
	DS			&Detector Suspension				\\
	GM			&Gifford-McMahon (cryocooler)	\\
	HEX		&Heat EXchanger					\\
	ILS		&Inner Lead Shield				\\
	IVC		&Inner Vacuum Chamber			\\
	FCS		&Fast Cooling System				\\
	LS			&Lead Suspension					\\
	MC			&Mixing Chamber					\\
	MSP		&Main Support Platform			\\ 
	NTD		&Neutron Transmutation Doped	\\
	OVC		&Outer Vacuum Chamber			\\
	PT			&Pulse Tube							\\
	TSP		&Tower Support Plate				\\
	WT			&Wire Tray
	
		\end{tabular}
	\end{flushleft}

%% file: main.bbl
\begin{thebibliography}{39}%
\makeatletter
\providecommand \@ifxundefined [1]{%
 \@ifx{#1\undefined}
}%
\providecommand \@ifnum [1]{%
 \ifnum #1\expandafter \@firstoftwo
 \else \expandafter \@secondoftwo
 \fi
}%
\providecommand \@ifx [1]{%
 \ifx #1\expandafter \@firstoftwo
 \else \expandafter \@secondoftwo
 \fi
}%
\providecommand \natexlab [1]{#1}%
\providecommand \enquote  [1]{``#1''}%
\providecommand \bibnamefont  [1]{#1}%
\providecommand \bibfnamefont [1]{#1}%
\providecommand \citenamefont [1]{#1}%
\providecommand \href@noop [0]{\@secondoftwo}%
\providecommand \href [0]{\begingroup \@sanitize@url \@href}%
\providecommand \@href[1]{\@@startlink{#1}\@@href}%
\providecommand \@@href[1]{\endgroup#1\@@endlink}%
\providecommand \@sanitize@url [0]{\catcode `\\12\catcode `\$12\catcode
  `\&12\catcode `\#12\catcode `\^12\catcode `\_12\catcode `\%12\relax}%
\providecommand \@@startlink[1]{}%
\providecommand \@@endlink[0]{}%
\providecommand \url  [0]{\begingroup\@sanitize@url \@url }%
\providecommand \@url [1]{\endgroup\@href {#1}{\urlprefix }}%
\providecommand \urlprefix  [0]{URL }%
\providecommand \Eprint [0]{\href }%
\providecommand \doibase [0]{http://dx.doi.org/}%
\providecommand \selectlanguage [0]{\@gobble}%
\providecommand \bibinfo  [0]{\@secondoftwo}%
\providecommand \bibfield  [0]{\@secondoftwo}%
\providecommand \translation [1]{[#1]}%
\providecommand \BibitemOpen [0]{}%
\providecommand \bibitemStop [0]{}%
\providecommand \bibitemNoStop [0]{.\EOS\space}%
\providecommand \EOS [0]{\spacefactor3000\relax}%
\providecommand \BibitemShut  [1]{\csname bibitem#1\endcsname}%
\let\auto@bib@innerbib\@empty
\bibitem [{\citenamefont {Alfonso}\ \emph {et~al.}(2015)\citenamefont {Alfonso}
  \emph {et~al.}}]{Alfonso:2015wka}%
  \BibitemOpen
  \bibfield  {author} {\bibinfo {author} {\bibfnamefont {K.}~\bibnamefont
  {Alfonso}} \emph {et~al.} (\bibinfo {collaboration} {CUORE Collaboration}),\
  }\href {\doibase 10.1103/PhysRevLett.115.102502} {\bibfield  {journal}
  {\bibinfo  {journal} {Phys.\ Rev.\ Lett.}\ }\textbf {\bibinfo {volume}
  {115}},\ \bibinfo {pages} {102502} (\bibinfo {year} {2015})}\BibitemShut
  {NoStop}%
\bibitem [{\citenamefont {Alduino}\ \emph {et~al.}(2018)\citenamefont {Alduino}
  \emph {et~al.}}]{Alduino:2017ehq}%
  \BibitemOpen
  \bibfield  {author} {\bibinfo {author} {\bibfnamefont {C.}~\bibnamefont
  {Alduino}} \emph {et~al.} (\bibinfo {collaboration} {CUORE Collaboration}),\
  }\href {\doibase 10.1103/PhysRevLett.120.132501} {\bibfield  {journal}
  {\bibinfo  {journal} {Phys.\ Rev.\ Lett.}\ }\textbf {\bibinfo {volume}
  {120}},\ \bibinfo {pages} {132501} (\bibinfo {year} {2018})}\BibitemShut
  {NoStop}%
\bibitem [{\citenamefont {Alduino}\ \emph {et~al.}(2016)\citenamefont {Alduino}
  \emph {et~al.}}]{Alduino:2016vjd}%
  \BibitemOpen
  \bibfield  {author} {\bibinfo {author} {\bibfnamefont {C.}~\bibnamefont
  {Alduino}} \emph {et~al.} (\bibinfo {collaboration} {CUORE Collaboration}),\
  }\href {\doibase 10.1088/1748-0221/11/07/P07009} {\bibfield  {journal}
  {\bibinfo  {journal} {J.\ Instrum.}\ }\textbf {\bibinfo {volume} {11}},\
  \bibinfo {pages} {P07009} (\bibinfo {year} {2016})}\BibitemShut {NoStop}%
\bibitem [{\citenamefont {Alduino}\ \emph
  {et~al.}(2017{\natexlab{a}})\citenamefont {Alduino} \emph
  {et~al.}}]{Alduino:2017pni}%
  \BibitemOpen
  \bibfield  {author} {\bibinfo {author} {\bibfnamefont {C.}~\bibnamefont
  {Alduino}} \emph {et~al.} (\bibinfo {collaboration} {CUORE Collaboration}),\
  }\href {\doibase 10.1140/epjc/s10052-017-5098-9} {\bibfield  {journal}
  {\bibinfo  {journal} {Eur.\ Phys.\ J.\ C}\ }\textbf {\bibinfo {volume}
  {77}},\ \bibinfo {pages} {532} (\bibinfo {year}
  {2017}{\natexlab{a}})}\BibitemShut {NoStop}%
\bibitem [{\citenamefont {Cushman}\ \emph {et~al.}(2017)\citenamefont {Cushman}
  \emph {et~al.}}]{Cushman:2016cnv}%
  \BibitemOpen
  \bibfield  {author} {\bibinfo {author} {\bibfnamefont {J.~S.}\ \bibnamefont
  {Cushman}} \emph {et~al.},\ }\href {\doibase 10.1016/j.nima.2016.11.020}
  {\bibfield  {journal} {\bibinfo  {journal} {Nucl.\ Instrum.\ Meth.\ A}\
  }\textbf {\bibinfo {volume} {844}},\ \bibinfo {pages} {32} (\bibinfo {year}
  {2017})}\BibitemShut {NoStop}%
\bibitem [{\citenamefont {Andreotti}\ \emph {et~al.}(2011)\citenamefont
  {Andreotti} \emph {et~al.}}]{Andreotti:2010vj}%
  \BibitemOpen
  \bibfield  {author} {\bibinfo {author} {\bibfnamefont {E.}~\bibnamefont
  {Andreotti}} \emph {et~al.},\ }\href {\doibase
  10.1016/j.astropartphys.2011.02.002} {\bibfield  {journal} {\bibinfo
  {journal} {Astropart.\ Phys.}\ }\textbf {\bibinfo {volume} {34}},\ \bibinfo
  {pages} {822} (\bibinfo {year} {2011})}\BibitemShut {NoStop}%
\bibitem [{\citenamefont {Arnaboldi}\ \emph {et~al.}(2008)\citenamefont
  {Arnaboldi} \emph {et~al.}}]{Arnaboldi:2008ds}%
  \BibitemOpen
  \bibfield  {author} {\bibinfo {author} {\bibfnamefont {C.}~\bibnamefont
  {Arnaboldi}} \emph {et~al.} (\bibinfo {collaboration} {Cuoricino
  Collaboration}),\ }\href {\doibase 10.1103/PhysRevC.78.035502} {\bibfield
  {journal} {\bibinfo  {journal} {Phys.\ Rev.\ C}\ }\textbf {\bibinfo {volume}
  {78}},\ \bibinfo {pages} {035502} (\bibinfo {year} {2008})}\BibitemShut
  {NoStop}%
\bibitem [{\citenamefont {Andreotti}\ \emph {et~al.}(2009)\citenamefont
  {Andreotti} \emph {et~al.}}]{Andreotti:2009zza}%
  \BibitemOpen
  \bibfield  {author} {\bibinfo {author} {\bibfnamefont {E.}~\bibnamefont
  {Andreotti}} \emph {et~al.},\ }\href {\doibase 10.1088/1748-0221/4/09/P09003}
  {\bibfield  {journal} {\bibinfo  {journal} {J.\ Instrum.}\ }\textbf {\bibinfo
  {volume} {4}},\ \bibinfo {pages} {P09003} (\bibinfo {year}
  {2009})}\BibitemShut {NoStop}%
\bibitem [{\citenamefont {Giachero}\ \emph {et~al.}(2013)\citenamefont
  {Giachero}, \citenamefont {Gotti}, \citenamefont {Maino},\ and\ \citenamefont
  {Pessina}}]{Giachero:2013iya}%
  \BibitemOpen
  \bibfield  {author} {\bibinfo {author} {\bibfnamefont {A.}~\bibnamefont
  {Giachero}}, \bibinfo {author} {\bibfnamefont {C.}~\bibnamefont {Gotti}},
  \bibinfo {author} {\bibfnamefont {M.}~\bibnamefont {Maino}}, \ and\ \bibinfo
  {author} {\bibfnamefont {G.}~\bibnamefont {Pessina}},\ }\href {\doibase
  10.1016/j.nima.2012.11.140} {\bibfield  {journal} {\bibinfo  {journal}
  {Nucl.\ Instrum.\ Meth.\ A}\ }\textbf {\bibinfo {volume} {718}},\ \bibinfo
  {pages} {229} (\bibinfo {year} {2013})}\BibitemShut {NoStop}%
\bibitem [{\citenamefont {Alduino}\ \emph
  {et~al.}(2017{\natexlab{b}})\citenamefont {Alduino} \emph
  {et~al.}}]{Alduino:2017qet}%
  \BibitemOpen
  \bibfield  {author} {\bibinfo {author} {\bibfnamefont {C.}~\bibnamefont
  {Alduino}} \emph {et~al.} (\bibinfo {collaboration} {CUORE Collaboration}),\
  }\href {\doibase 10.1140/epjc/s10052-017-5080-6} {\bibfield  {journal}
  {\bibinfo  {journal} {Eur.\ Phys.\ J.\ C}\ }\textbf {\bibinfo {volume}
  {77}},\ \bibinfo {pages} {543} (\bibinfo {year}
  {2017}{\natexlab{b}})}\BibitemShut {NoStop}%
\bibitem [{\citenamefont {Ardito}\ and\ \citenamefont
  {Perotti}(2018)}]{Ardito:2011SA}%
  \BibitemOpen
  \bibfield  {author} {\bibinfo {author} {\bibfnamefont {R.}~\bibnamefont
  {Ardito}}\ and\ \bibinfo {author} {\bibfnamefont {F.}~\bibnamefont
  {Perotti}},\ }\href@noop {} {\enquote {\bibinfo {title} {{Seismic analysis of
  support structure for the CUORE experiment cryostat (in Italian)}},}\ }
  (\bibinfo {year} {2011~and~2018}),\ \bibinfo {note} {{Report of the
  Polytechnic University of Milan}}\BibitemShut {NoStop}%
\bibitem [{\citenamefont {Bucci}\ \emph {et~al.}(2017)\citenamefont {Bucci},
  \citenamefont {Carniti}, \citenamefont {Cassina}, \citenamefont {Gotti},
  \citenamefont {Pelosi}, \citenamefont {Pessina}, \citenamefont {Turqueti},\
  and\ \citenamefont {Zimmermann}}]{Bucci:2017gew}%
  \BibitemOpen
  \bibfield  {author} {\bibinfo {author} {\bibfnamefont {C.}~\bibnamefont
  {Bucci}}, \bibinfo {author} {\bibfnamefont {P.}~\bibnamefont {Carniti}},
  \bibinfo {author} {\bibfnamefont {L.}~\bibnamefont {Cassina}}, \bibinfo
  {author} {\bibfnamefont {C.}~\bibnamefont {Gotti}}, \bibinfo {author}
  {\bibfnamefont {A.}~\bibnamefont {Pelosi}}, \bibinfo {author} {\bibfnamefont
  {G.}~\bibnamefont {Pessina}}, \bibinfo {author} {\bibfnamefont
  {M.}~\bibnamefont {Turqueti}}, \ and\ \bibinfo {author} {\bibfnamefont
  {S.}~\bibnamefont {Zimmermann}},\ }\href {\doibase
  10.1088/1748-0221/12/12/P12013} {\bibfield  {journal} {\bibinfo  {journal}
  {J.\ Instrum.}\ }\textbf {\bibinfo {volume} {12}},\ \bibinfo {pages} {P12013}
  (\bibinfo {year} {2017})}\BibitemShut {NoStop}%
\bibitem [{\citenamefont {Barucci}\ \emph {et~al.}(2008)\citenamefont
  {Barucci}, \citenamefont {Lolli}, \citenamefont {Risegari},\ and\
  \citenamefont {Ventura}}]{Barucci:2008zz}%
  \BibitemOpen
  \bibfield  {author} {\bibinfo {author} {\bibfnamefont {M.}~\bibnamefont
  {Barucci}}, \bibinfo {author} {\bibfnamefont {L.}~\bibnamefont {Lolli}},
  \bibinfo {author} {\bibfnamefont {L.}~\bibnamefont {Risegari}}, \ and\
  \bibinfo {author} {\bibfnamefont {G.}~\bibnamefont {Ventura}},\ }\href
  {\doibase 10.1016/j.cryogenics.2008.03.018} {\bibfield  {journal} {\bibinfo
  {journal} {Cryogenics}\ }\textbf {\bibinfo {volume} {48}},\ \bibinfo {pages}
  {166} (\bibinfo {year} {2008})}\BibitemShut {NoStop}%
\bibitem [{\citenamefont {Vanzini}\ \emph {et~al.}(2001)\citenamefont {Vanzini}
  \emph {et~al.}}]{Vanzini:2001rx}%
  \BibitemOpen
  \bibfield  {author} {\bibinfo {author} {\bibfnamefont {M.}~\bibnamefont
  {Vanzini}} \emph {et~al.},\ }\href {\doibase 10.1016/S0168-9002(00)01228-6}
  {\bibfield  {journal} {\bibinfo  {journal} {Nucl.\ Instrum.\ Meth.\ A}\
  }\textbf {\bibinfo {volume} {461}},\ \bibinfo {pages} {293} (\bibinfo {year}
  {2001})}\BibitemShut {NoStop}%
\bibitem [{\citenamefont {D'Addabbo}\ \emph {et~al.}(2018)\citenamefont
  {D'Addabbo}, \citenamefont {Bucci}, \citenamefont {Canonica}, \citenamefont
  {Di~Domizio}, \citenamefont {Gorla}, \citenamefont {Marini}, \citenamefont
  {Nucciotti}, \citenamefont {Nutini}, \citenamefont {Rusconi},\ and\
  \citenamefont {Welliver}}]{DAddabbo:2017efe}%
  \BibitemOpen
  \bibfield  {author} {\bibinfo {author} {\bibfnamefont {A.}~\bibnamefont
  {D'Addabbo}}, \bibinfo {author} {\bibfnamefont {C.}~\bibnamefont {Bucci}},
  \bibinfo {author} {\bibfnamefont {L.}~\bibnamefont {Canonica}}, \bibinfo
  {author} {\bibfnamefont {S.}~\bibnamefont {Di~Domizio}}, \bibinfo {author}
  {\bibfnamefont {P.}~\bibnamefont {Gorla}}, \bibinfo {author} {\bibfnamefont
  {L.}~\bibnamefont {Marini}}, \bibinfo {author} {\bibfnamefont
  {A.}~\bibnamefont {Nucciotti}}, \bibinfo {author} {\bibfnamefont
  {I.}~\bibnamefont {Nutini}}, \bibinfo {author} {\bibfnamefont
  {C.}~\bibnamefont {Rusconi}}, \ and\ \bibinfo {author} {\bibfnamefont
  {B.}~\bibnamefont {Welliver}},\ }\href {\doibase
  10.1016/j.cryogenics.2018.05.001} {\bibfield  {journal} {\bibinfo  {journal}
  {Cryogenics}\ }\textbf {\bibinfo {volume} {93}},\ \bibinfo {pages} {56}
  (\bibinfo {year} {2018})}\BibitemShut {NoStop}%
\bibitem [{Min()}]{MinusK}%
  \BibitemOpen
  \href {https://www.minusk.com/} {{\bibinfo {title}
  {{https://www.minusk.com}}}}\BibitemShut {NoStop}%
\bibitem [{\citenamefont {Bersani}\ \emph {et~al.}(2019)\citenamefont {Bersani}
  \emph {et~al.}}]{Bersani:DS_prep}%
  \BibitemOpen
  \bibfield  {author} {\bibinfo {author} {\bibfnamefont {A.}~\bibnamefont
  {Bersani}} \emph {et~al.},\ }\href@noop {} {\enquote {\bibinfo {title} {{The
  CUORE Detector Suspension}},}\ } (\bibinfo {year} {2019})\ \bibinfo {note}
  {[paper in preparation]}\BibitemShut {NoStop}%
\bibitem [{\citenamefont {Bersani}\ \emph {et~al.}(2013)\citenamefont
  {Bersani}, \citenamefont {Canonica}, \citenamefont {Cariello}, \citenamefont
  {Cereseto}, \citenamefont {Di~Domizio},\ and\ \citenamefont
  {Pallavicini}}]{Bersani:2013zta}%
  \BibitemOpen
  \bibfield  {author} {\bibinfo {author} {\bibfnamefont {A.}~\bibnamefont
  {Bersani}}, \bibinfo {author} {\bibfnamefont {L.}~\bibnamefont {Canonica}},
  \bibinfo {author} {\bibfnamefont {M.}~\bibnamefont {Cariello}}, \bibinfo
  {author} {\bibfnamefont {R.}~\bibnamefont {Cereseto}}, \bibinfo {author}
  {\bibfnamefont {S.}~\bibnamefont {Di~Domizio}}, \ and\ \bibinfo {author}
  {\bibfnamefont {M.}~\bibnamefont {Pallavicini}},\ }\href {\doibase
  10.1016/j.cryogenics.2012.10.005} {\bibfield  {journal} {\bibinfo  {journal}
  {Cryogenics}\ }\textbf {\bibinfo {volume} {54}},\ \bibinfo {pages} {50}
  (\bibinfo {year} {2013})}\BibitemShut {NoStop}%
\bibitem [{\citenamefont {Alessandria}\ \emph {et~al.}(2013)\citenamefont
  {Alessandria} \emph {et~al.}}]{Alessandria:2013ufa}%
  \BibitemOpen
  \bibfield  {author} {\bibinfo {author} {\bibfnamefont {F.}~\bibnamefont
  {Alessandria}} \emph {et~al.},\ }\href {\doibase 10.1016/j.nima.2013.06.015}
  {\bibfield  {journal} {\bibinfo  {journal} {Nucl.\ Instrum.\ Meth.\ A}\
  }\textbf {\bibinfo {volume} {727}},\ \bibinfo {pages} {65} (\bibinfo {year}
  {2013})}\BibitemShut {NoStop}%
\bibitem [{CuN()}]{CuNOSV_aurubis}%
  \BibitemOpen
  \href {https://www.aurubis.com/products/page-shapes/} {{\bibinfo
  {title} {{https://www.aurubis.com/products/page-shapes}}}}\BibitemShut
  {NoStop}%
\bibitem [{\citenamefont {Schwark}\ \emph {et~al.}(1983)\citenamefont
  {Schwark}, \citenamefont {Pobell}, \citenamefont {Halperin}, \citenamefont
  {Buchal}, \citenamefont {Hanssen}, \citenamefont {Kubota},\ and\
  \citenamefont {Mueller}}]{Schwark&al:1983}%
  \BibitemOpen
  \bibfield  {author} {\bibinfo {author} {\bibfnamefont {M.}~\bibnamefont
  {Schwark}}, \bibinfo {author} {\bibfnamefont {F.}~\bibnamefont {Pobell}},
  \bibinfo {author} {\bibfnamefont {W.~P.}\ \bibnamefont {Halperin}}, \bibinfo
  {author} {\bibfnamefont {C.}~\bibnamefont {Buchal}}, \bibinfo {author}
  {\bibfnamefont {J.}~\bibnamefont {Hanssen}}, \bibinfo {author} {\bibfnamefont
  {M.}~\bibnamefont {Kubota}}, \ and\ \bibinfo {author} {\bibfnamefont {R.~M.}\
  \bibnamefont {Mueller}},\ }\href {\doibase 10.1007/BF00684000} {\bibfield
  {journal} {\bibinfo  {journal} {J.\ Low Temp.\ Phys.}\ }\textbf {\bibinfo
  {volume} {53}},\ \bibinfo {pages} {685} (\bibinfo {year} {1983})}\BibitemShut
  {NoStop}%
\bibitem [{\citenamefont {Kol\'a\u{c}}\ \emph {et~al.}(1985)\citenamefont
  {Kol\'a\u{c}}, \citenamefont {Neganov},\ and\ \citenamefont
  {Sahling}}]{Kolac&al:1984}%
  \BibitemOpen
  \bibfield  {author} {\bibinfo {author} {\bibfnamefont {M.}~\bibnamefont
  {Kol\'a\u{c}}}, \bibinfo {author} {\bibfnamefont {B.~S.}\ \bibnamefont
  {Neganov}}, \ and\ \bibinfo {author} {\bibfnamefont {S.}~\bibnamefont
  {Sahling}},\ }\href {\doibase 10.1007/BF00682449} {\bibfield  {journal}
  {\bibinfo  {journal} {J.\ Low Temp.\ Phys.}\ }\textbf {\bibinfo {volume}
  {59}},\ \bibinfo {pages} {547} (\bibinfo {year} {1985})}\BibitemShut
  {NoStop}%
\bibitem [{\citenamefont {Martinez}\ \emph {et~al.}(2009)\citenamefont
  {Martinez} \emph {et~al.}}]{Martinez:2009gjv}%
  \BibitemOpen
  \bibfield  {author} {\bibinfo {author} {\bibfnamefont {M.}~\bibnamefont
  {Martinez}} \emph {et~al.},\ }\href {\doibase 10.1063/1.3292436} {\bibfield
  {journal} {\bibinfo  {journal} {AIP Conf. Proc.}\ }\textbf {\bibinfo {volume}
  {1185}},\ \bibinfo {pages} {693} (\bibinfo {year} {2009})}\BibitemShut
  {NoStop}%
\bibitem [{\citenamefont {Bucci}\ \emph
  {et~al.}(2019{\natexlab{a}})\citenamefont {Bucci} \emph
  {et~al.}}]{Bucci:ILS_prep}%
  \BibitemOpen
  \bibfield  {author} {\bibinfo {author} {\bibfnamefont {C.}~\bibnamefont
  {Bucci}} \emph {et~al.},\ }\href@noop {} {\enquote {\bibinfo {title} {{The
  Roman lead shield of the CUORE experiment}},}\ } (\bibinfo {year}
  {2019}{\natexlab{a}})\ \bibinfo {note} {[paper in preparation]}\BibitemShut
  {NoStop}%
\bibitem [{\citenamefont {Alessandrello}\ \emph {et~al.}(1998)\citenamefont
  {Alessandrello} \emph {et~al.}}]{Alessandrello:1998RL}%
  \BibitemOpen
  \bibfield  {author} {\bibinfo {author} {\bibfnamefont {A.}~\bibnamefont
  {Alessandrello}} \emph {et~al.},\ }\href {\doibase
  10.1016/S0168-583X(98)00279-1} {\bibfield  {journal} {\bibinfo  {journal}
  {Nucl.\ Instrum.\ Meth.\ B}\ }\textbf {\bibinfo {volume} {142}},\ \bibinfo
  {pages} {163} (\bibinfo {year} {1998})}\BibitemShut {NoStop}%
\bibitem [{\citenamefont {{American Society of Mechanical Engineers
  (ASME)}}(2010)}]{ASME-BPVC}%
  \BibitemOpen
  \bibfield  {author} {\bibinfo {author} {\bibnamefont {{American Society of
  Mechanical Engineers (ASME)}}},\ }\href
  {https://www.asme.org/products/codes-standards/bpvcviii1-2015-bpvc-section-viiirules}
  {\emph {\bibinfo {title} {{Boiler \& Pressure Vessel Code}}}} (\bibinfo
  {year} {2010}),\ \bibinfo {note} {{Sec.\ VIII, Div.~1}}\BibitemShut
  {NoStop}%
\bibitem [{ANS()}]{ANSYS-web}%
  \BibitemOpen
  \href {https://www.ansys.com/} {{\bibinfo {title}
  {{https://www.ansys.com}}}}\BibitemShut {NoStop}%
\bibitem [{Hel()}]{Helicoflex_technetics}%
  \BibitemOpen
  \href
  {https://technetics.com/products/sealing-solutions/metal-seals/helicoflex/}
  {{\bibinfo {title}
  {{https://technetics.com/products/sealing-solutions/metal-seals/helicoflex}}}}\BibitemShut {NoStop}%
\bibitem [{\citenamefont {Arblaster}(2015)}]{Arblaster:2015Cu}%
  \BibitemOpen
  \bibfield  {author} {\bibinfo {author} {\bibfnamefont {J.~W.}\ \bibnamefont
  {Arblaster}},\ }\href {\doibase 10.1007/s11669-015-0399-x} {\bibfield
  {journal} {\bibinfo  {journal} {J.\ Phase Equil.\ Diff.}\ }\textbf {\bibinfo
  {volume} {36}},\ \bibinfo {pages} {422} (\bibinfo {year} {2015})}\BibitemShut
  {NoStop}%
\bibitem [{\citenamefont {Arblaster}(2012)}]{Arblaster:2012Pb}%
  \BibitemOpen
  \bibfield  {author} {\bibinfo {author} {\bibfnamefont {J.~W.}\ \bibnamefont
  {Arblaster}},\ }\href {\doibase 10.1016/j.calphad.2012.08.004} {\bibfield
  {journal} {\bibinfo  {journal} {Comput.\ Coupling Phase Diagrams
  Thermochem.}\ }\textbf {\bibinfo {volume} {39}},\ \bibinfo {pages} {47}
  (\bibinfo {year} {2012})}\BibitemShut {NoStop}%
\bibitem [{\citenamefont {White}\ \emph {et~al.}(1990)\citenamefont {White},
  \citenamefont {Collocott},\ and\ \citenamefont {Collins}}]{White:1990}%
  \BibitemOpen
  \bibfield  {author} {\bibinfo {author} {\bibfnamefont {G.~K.}\ \bibnamefont
  {White}}, \bibinfo {author} {\bibfnamefont {S.}~\bibnamefont {Collocott}}, \
  and\ \bibinfo {author} {\bibfnamefont {J.~G.}\ \bibnamefont {Collins}},\
  }\href {http://stacks.iop.org/0953-8984/2/i=37/a=015} {\bibfield  {journal}
  {\bibinfo  {journal} {J.\ Phys. Condensed Matter}\ }\textbf {\bibinfo
  {volume} {2}},\ \bibinfo {pages} {7715} (\bibinfo {year} {1990})}\BibitemShut
  {NoStop}%
\bibitem [{\citenamefont {Pagliarone}\ \emph {et~al.}(2018)\citenamefont
  {Pagliarone}, \citenamefont {Cappelli}, \citenamefont {Bucci}, \citenamefont
  {Gorla}, \citenamefont {D'Aguanno}, \citenamefont {Marignetti}, \citenamefont
  {Erme},\ and\ \citenamefont {Kartal}}]{Pagliarone:2018mpj}%
  \BibitemOpen
  \bibfield  {author} {\bibinfo {author} {\bibfnamefont {C.}~\bibnamefont
  {Pagliarone}}, \bibinfo {author} {\bibfnamefont {L.}~\bibnamefont
  {Cappelli}}, \bibinfo {author} {\bibfnamefont {C.}~\bibnamefont {Bucci}},
  \bibinfo {author} {\bibfnamefont {P.}~\bibnamefont {Gorla}}, \bibinfo
  {author} {\bibfnamefont {D.}~\bibnamefont {D'Aguanno}}, \bibinfo {author}
  {\bibfnamefont {F.}~\bibnamefont {Marignetti}}, \bibinfo {author}
  {\bibfnamefont {G.}~\bibnamefont {Erme}}, \ and\ \bibinfo {author}
  {\bibfnamefont {S.}~\bibnamefont {Kartal}},\ }\href {\doibase
  10.22323/1.314.0634} {\bibfield  {journal} {\bibinfo  {journal} {PoS}\
  }\textbf {\bibinfo {volume} {EPS-HEP2017}},\ \bibinfo {pages} {634} (\bibinfo
  {year} {2018})}\BibitemShut {NoStop}%
\bibitem [{\citenamefont {Bucci}\ \emph
  {et~al.}(2019{\natexlab{b}})\citenamefont {Bucci} \emph
  {et~al.}}]{Bucci:FCS_prep}%
  \BibitemOpen
  \bibfield  {author} {\bibinfo {author} {\bibfnamefont {C.}~\bibnamefont
  {Bucci}} \emph {et~al.},\ }\href@noop {} {\enquote {\bibinfo {title} {{CUORE
  Fast Cooling}},}\ } (\bibinfo {year} {2019}{\natexlab{b}})\ \bibinfo {note}
  {[paper in preparation]}\BibitemShut {NoStop}%
\bibitem [{Cry({\natexlab{a}})}]{Cryomech_AL600}%
  \BibitemOpen
  \href {http://www.cryomech.com/cryorefrigerators/gifford/al600/} {
  {\bibinfo {title}
  {{http://www.cryomech.com/cryorefrigerators/gifford/al600}}}} \BibitemShut {NoStop}%
\bibitem [{Cry({\natexlab{b}})}]{Cryomech_PT415}%
  \BibitemOpen
  \href {http://www.cryomech.com/cryorefrigerators/pulse-tube/pt415/} {
  {\bibinfo {title}
  {{http://www.cryomech.com/cryorefrigerators/pulse-tube/pt415/}}}}
  \BibitemShut {NoStop}%
\bibitem [{LC_()}]{LC_CF}%
  \BibitemOpen
  \href {https://leiden-cryogenics.com/products/cryogen-free/} {
  {\bibinfo {title} {{https://leiden-cryogenics.com/products/cryogen-free}}}
  }\BibitemShut {NoStop}%
\bibitem [{MFF()}]{MFFT}%
  \BibitemOpen
  \href {http://www.magnicon.com/squid-systems/noise-thermometer/} {
  {\bibinfo {title}
  {{http://www.magnicon.com/squid-systems/noise-thermometer}}}}\BibitemShut
  {NoStop}%
\bibitem [{\citenamefont {{S.\ Dell'Oro}}(2019)}]{Delloro_PhD-thesis:2017}%
  \BibitemOpen
  \bibfield  {author} {\bibinfo {author} {\bibnamefont {{S. Dell'Oro}}},\ }\href
  {https://cuore.lngs.infn.it/en/publications/theses}
  {\emph {\bibinfo {title} {{Ph.\,D.\ thesis, INFN - Gran
  Sasso Science Institute}}}} (\bibinfo
  {year} {2019})\BibitemShut
  {NoStop}%
\bibitem [{\citenamefont {Benato}\ \emph {et~al.}(2018)\citenamefont {Benato}
  \emph {et~al.}}]{Benato:2017kdf}%
  \BibitemOpen
  \bibfield  {author} {\bibinfo {author} {\bibfnamefont {G.}~\bibnamefont
  {Benato}} \emph {et~al.},\ }\href {\doibase 10.1088/1748-0221/13/01/P01010}
  {\bibfield  {journal} {\bibinfo  {journal} {J.\ Instrum.}\ }\textbf {\bibinfo
  {volume} {13}},\ \bibinfo {pages} {P01010} (\bibinfo {year}
  {2018})}\BibitemShut {NoStop}%
\end{thebibliography}%
